\newif\ifsubmode
\submodefalse

\ifsubmode
\documentclass[10pt,preprint]{aastex}
  \received{}
  \revised{}
  \accepted{}
\else
  \documentclass[iop]{emulateapj}   \usepackage{apjfonts}
\fi

\usepackage{natbib,longtable}
\usepackage{epsf}
\usepackage{color}
\usepackage{amsmath}
\ifsubmode
   
\else
  
\fi

\newcommand{\hst}{\textit{HST}}

\newcommand{\spitzer}{\textit{Spitzer}}

\newcommand{\lsim}{\lesssim}
\newcommand{\gsim}{\gtrsim}

\newcommand{\eg}{e.g.}

\newcommand{\msol}{\hbox{$M_\odot$}}

\newcommand{\mone}{\hbox{$[3.6]$}}

\newcommand{\reff}{\hbox{$r_\mathrm{eff}$}}

\newcommand{\rproj}{\hbox{$R_\mathrm{proj}$}}
\newcommand{\bs}{\hbox{$\!\!\!\!$}}
\newcommand{\sersic}{S\'ersic}
\newcommand{\Sersic}{S\'ersic}

\ifsubmode
\else
\renewcommand{\bs}{\hbox{}}
\fi

\shorttitle{The Environmental Dependence of the Color-Mass-Morphology Relation at $z=1.6$}
\shortauthors{Bassett et al.}

\begin{document}
\title{CANDELS OBSERVATIONS OF THE ENVIRONMENTAL DEPENDENCE OF THE
  \\ COLOR--MASS--MORPHOLOGY RELATION AT  $Z=1.6$}
\author{Robert Bassett\altaffilmark{1,2,9}, 
Casey  Papovich\altaffilmark{1,2}, 
Jennifer M. Lotz\altaffilmark{3}, 
Eric F. Bell\altaffilmark{4},
Steven  L. Finkelstein\altaffilmark{5}, 
Jeffrey~A.~Newman\altaffilmark{6},
Kim-Vy Tran\altaffilmark{1,2}, 
Omar Almaini\altaffilmark{7},
Caterina Lani\altaffilmark{7},
Michael Cooper\altaffilmark{8},
Darren Croton\altaffilmark{9},
Avishai~Dekel\altaffilmark{10},
Henry~C.~Ferguson\altaffilmark{3},
Dale~D.~Kocevski\altaffilmark{11},
Anton~M.~Koekemoer\altaffilmark{3}, 
David~C.~Koo\altaffilmark{11}, 
Elizabeth~J.~McGrath\altaffilmark{11},
Daniel~H.~McIntosh\altaffilmark{12}, 
Risa Wechsler\altaffilmark{13}}
\affil{$^1$George P.\ and Cynthia Woods Mitchell Institute for
  Fundamental Physics and Astronomy, College Station, TX, 77843-4242}
\affil{$^2$Department of Physics and Astronomy, Texas A\&M University, College
Station, TX, 77843-4242}
\affil{$^3$Space Telescope Science Institute, 3700 San Martin
  Dr., Baltimore, MD 21218}
\affil{$^4$Department of Astronomy, University of Michigan,  Ann
  Arbor, MI 48109}
\affil{$^5$Hubble Fellow, Department of Astronomy, University of Texas,
  Austin, TX 78712}
\affil{$^{6}$University of Pittsburgh, Department of Physics and
Astronomy, Pittsburgh, PA 15260}
\affil{$^7$School of Physics \& Astronomy, University of
  Nottingham, Nottingham, UK}
\affil{$^8$Hubble Fellow, Center for Galaxy Evolution, Department of
  Physics and Astronomy, University of California, Irvine, 4129 Frederick
 Reines Hall, Irvine, CA 92697, USA}
\affil{$^9$Centre for Astrophysics and Supercomputing, Swinburne
  University of Technology, Hawthorn, Australia}
\affil{$^{10}$Racah Institute of Physics, The Hebrew University, Jerusalem 91904, Israel}
\affil{$^{11}$UCO/ Lick Observatory, Department of Astronomy and
  Astrophysics, University of California, Santa Cruz, CA 95064}
\affil{$^{12}$Department of Physics, University of Missouri-Kansas City, 5110 Rockhill Road, Kansas City, MO 64110}
\affil{$^{13}$Kavli Institute for Particle Astrophysics and Cosmology,
  Physics Department, and SLAC National Accelerator Laboratory, Stanford University, Stanford, CA 94305}

\begin{abstract} 
\noindent
We study the environmental dependence of color, stellar mass, and
morphology 
by comparing galaxies in a forming cluster 
to those in the field at
$z=1.6$ with \textit{Hubble Space Telescope} near-infrared imaging in
the CANDELS/UDS field.  We quantify the morphology of the galaxies
using the effective radius, $\reff$,  and \sersic\ index, $n$.   In
both the cluster and field, approximately half of the bulge-dominated
galaxies ($n>2$) reside on the red sequence of the color-magnitude diagram, and most
disk-dominated galaxies ($n<2$) have colors expected for star-forming
galaxies. There is weak evidence that cluster
galaxies have redder rest-frame $U-B$ colors and higher stellar masses
compared to the field.  Star-forming galaxies in both the cluster and
field show no significant differences in their morphologies.  In
contrast, there is evidence that quiescent galaxies in the cluster
have larger median effective radii
and smaller \sersic\ indices 
compared to the field with a significance of $2\sigma$.
These differences are most pronounced for galaxies at
clustercentric distances 1~Mpc~$<\rproj<$~1.5~Mpc, which have 
low \sersic\ indices and possibly larger effective radii, 
more consistent with star-forming galaxies at this
  epoch and in contrast to other quiescent galaxies.
We argue that star-forming galaxies are processed under the influence
of the cluster environment at distances greater than the cluster-halo
virial radius.  Our results are consistent with models where gas
accretion onto these galaxies is suppressed from processes
associated with the cluster environment.   

\end{abstract}


\section{INTRODUCTION}

Galaxies in the low redshift Universe show a very strong dependence
between environment, morphology, color, and stellar mass
\citep[\eg,][]{hogg04,kauf04,skibba09}.   Galaxies in regions of
higher local density, such as galaxy groups and
clusters, have lower levels of star-formation, higher fractions of
red, passive galaxies, and more early-type morphologies compared to
galaxies in lower local number density regions
\citep[\eg,][]{dres80,post84}. \citet[and others]{poggi08} find
that this relation extends to intermediate redshifts ($z\sim 0.4-0.8$)
with the highest density regions being exclusively populated by
elliptical morphological types.

It is difficult, however, to separate the galaxy properties 
which depend on high local density (regions of dense ``environment'')
from those which depend on galaxy mass.  One challenge is that
galaxies in regions of high local density have
higher average halo and stellar masses, and care must be taken to
study environmental effects on samples at fixed stellar mass. For example,
\citet{kauf04} showed that the red, passive galaxies at $z\sim 0.1$
that dominate in high local density regions
have highly concentrated morphologies, but at fixed mass their sizes
and concentrations are almost independent of environment.
\citet{guo09} find that the concentrations of the central galaxies of
dark matter halos at $0 \lsim z \leq 0.2$ depends strongly on their
stellar mass, but only weakly on their halo mass, implying a weak
dependence on environment.  Similarly, \citet{vanderwel08b} show that
galaxies at $z\sim 0.02$ with the highest masses show no dependence
between morphological concentration and environment (local density of galaxies).  It is
only for lower mass galaxies ($M \sim 3\times 10^{10} - 5 \times
10^{10}$~\msol) that there is a weak trend toward higher galaxy
concentration in higher local density regions.
Therefore, the environment has a subtle effect on
galaxy morphologies at low redshifts. 

\citet{grutz11a} show that these trends persist to
higher redshifts ($z\sim 0.4-1.0$) and that the intrinsic properties
of a galaxy are more dependent on stellar mass than local
density at earlier epochs.  In a similar redshift range some studies
attempted to disentangle the color-mass-density relation and its
evolution by examining the colors or morphologies of galaxies as a
function of local density in narrow bins of redshift and
stellar mass \citep{tasca09,cucci10}. These studies found that the
correlation between color, mass, and density is considerably weaker 
at higher redshifts with the exception of
galaxies in the low mass regime. The implication is that these lower 
mass galaxies were formed more recently in a
time when evolved large-scale structures were already in place.

In contrast, the environment does appear to have a strong effect on
the relative star-formation rates of galaxies.   \citet{hogg03} showed
that at $z\sim 0.2$ the density of galaxies correlates
with color: redder galaxies lie in regions of higher
local density   \citep[see
also,][]{blan05b,baldry06}.  Similarly, \citet{kauf04} found that
higher density regions anticorrelate most strongly with galaxy
specific SFR (sSFR, the SFR per unit stellar mass):  galaxies in
high local number density regions have lower average
sSFRs.  \citet{patel09} show that this trend also
persists to higher redshifts ($z\sim 0.7$). Interestingly, most of
this effect on the sSFRs does not arise from an environmental
dependence on the properties of star-forming galaxies, which  appear
unchanged  between regions of high and low
local number density
\citep{hansen09,peng10,wetzel11b}.   Instead, it is the fraction of
quenched galaxies that increases with increasing galaxy density that
drives this observation \citep{peng10}.    

Several authors have argued that this quenching fraction is different
for galaxies that are ``centrals'' or ``satellites'' in their halos,
where quenching in centrals depends primarily on the galaxy's halo
mass while the quenched fraction of galaxies that are ``satellites''
depends on their distance from the halo center
\citep[\eg,][]{wetzel11b,wilman11,woo12}.  \citet{hansen09} find that
the fraction of red (quenched) and blue (star-forming) galaxies
changes dramatically with distance from the center of clusters.
\citet{wetzel11b} find that the fraction of quenched galaxies persists
out to $\approx 2 R_\mathrm{vir}$, where $R_\mathrm{vir}$ is the
virial radius of the halo \citep[\eg,][]{hansen09}. Selecting galaxies in bins of
morphological type, \citet{guo09} find no evidence that the structures
of early-type galaxies differ between central and satellite galaxies
matched in both optical color and stellar mass, although
\citet{wein09} find that  late-type satellite galaxies show smaller
radii and larger concentrations than late-type centrals.
\citet{wein09} also find that late-type satellites have redder colors
than late-type central galaxies.    This suggests that the stellar
mass of the galaxy is the more fundamental property determining the
structure of early-type galaxies, and that  observed differences in
satellite and central galaxies may be due to the quenching of galaxies
after they become satellites.
\citet{woo12} argue these observations are consistent with theoretical
expectations that quenching mechanisms affect the satellites as they
move into the environment of the larger mass halo \citep[\eg,][]{bahe12}. 

It is unclear how these trends between color, SFR, morphology, mass
and environment evolve with redshift.   Both \citet{cooper07} and
\citet{elbaz07} observed that the low redshift SFR-density relation
evolves and possibly reverses around $z \sim 1$.   It has also been
observed that at $z\sim 1-1.6$ the surface density of IR-luminous,
star-forming galaxies increases in regions of high local density
\citep[\eg,][]{tran10,koce11}.  However, the study of
\citet{grutz11b} found that in the redshift range $1 \lsim z \lsim 3$
there is a strong dependence between SFR (and sSFR) and stellar mass,
and very weak (if any) dependence between SFR and environment
\citep[see also,][]{bauer11}.
It is only in the highest local number density regions where star formation is
significantly suppressed, and this is only apparent to $z\sim 2$.  However,
studies show that even at these redshifts, regions of high local
density contain a dominant population of mostly quiescent galaxies,
and the overall trend in color and specific SFR mostly unchanged from
$z\sim 0$ \citep{cooper08,chuter11,quadri12}.

Other studies have looked at the relation between galaxy size
(morphology) and environment at $z > 1$ \citep{cooper11,
zirm11,papo12}.  These studies have found that  at fixed mass, spheroidal
(bulge-dominated, with \sersic\ index $n > 2$)  galaxies in regions of
higher density have larger sizes compared to spheroidal galaxies in
low-density regions.  This seems to imply that there are processes
associated with the higher density regions that enhance or accelerate
galaxy growth.  One obvious candidate is an apparent higher merger rate for
galaxies in higher density environments.  Indeed,  \citet{lotz11}
observe an enhancement in the rate of dissipationless (gas poor, or
``dry'') mergers in galaxies in a forming cluster at $z=1.62$ compared
to similarly selected galaxies in the field \citep[similar to the
findings of][at $z=0.83$]{tran05}.   If these ``dry''  mergers are in
fact the dominant driver of the size evolution of early type galaxies,
they may also affect the galaxy morphologies \citep[\eg,][]{nava90},
and we may expect to find variation in galaxy  structure at this
epoch.  

One reason it is difficult to
disentangle galaxy evolution effects arising from ``environment''
from those which depend on ``mass'', is that different methods are used
to define ``environment'' and these are sensitive to the effects of
environments on different scales. Many methods have appeared in the literature such as fixed aperture
and nearest neighbor distances, but it has been shown that there is
considerable scatter between them. \citet{muldrew12} compare and
contrast different environment estimators  using mock galaxy samples
taken from the Millennium simulation.    They found that different
methods probe different aspects of the dark matter halos that contain
groups and clusters.   For example, nearest  neighbor statistics
better study the internal halo properties and effects of smaller
structures such as group sized halos. The fixed aperture method, on
the other hand, is found to be a better probe of the halo as a whole.
However, all of these environment statistics are sensitive to errors
in galaxy distances (in particular redshift errors), which can be the
limiting factors in this analysis. 

Here, we extend these studies to understand how galaxy color, stellar
mass, and  morphology depend on  environment at $z=1.6$  using a
portion of the data from the UKIRT Infrared Deep Sky
Survey--Ultra-Deep Survey (UKIDSS-UDS) field covered with the Hubble
Space Telescope (\hst) as part of the Cosmic Assembly Near-infrared
Deep Extragalactic Legacy Survey \citep[CANDELS,][]{grogin11,koek11}.
These data cover one of the largest contiguous fields with WFC3
near-IR imaging.  At $z=1.6$ this field covers both a forming cluster
at $z=1.62$ and galaxies in the field.  Therefore, we are able to study galaxies
within a single redshift slice in this large field. This allows us to test
effects associated with environment over a very large range in local
galaxy density,
from the rich environment of a forming cluster to the low density field using a
highly homogeneous dataset. Furthermore, because we study galaxies
from a single dataset with uniform analysis, any bias from systematics
affects all galaxies in a similar way. This means we can make robust
(relative) comparisons between galaxies in different environments. 

Throughout this paper we report magnitudes measured relative to the AB
system \citep{oke83} and we assume cosmological parameters $\Omega_m =
0.3$, $\Omega_\Lambda = 0.7$, and $H_0 = 70$ km s$^{-1}$ Mpc$^{-1}$,
which give an angular diameter distance of 0.508~Mpc arcmin$^{-1}$
(8.47~kpc arcsec$^{-1}$) for $z=1.6$.  All projected distances are
given in physical (proper) distances at $z=1.6$. 

\section{DATA}\label{section:data}

For this paper, we use available datasets that cover
the CANDELS UDS field.  The available data in this field includes \hst/WFC3 imaging in the
F125W and F160W filters from CANDELS over  $9\farcm4 \times 22\farcm0$
\citep{grogin11,koek11}, corresponding to a projected,
physical area 4.8~Mpc $\times 11.2$~Mpc at $z=1.6$. 

Here we used  a photometric catalog containing
photometry from the UKIDSS-UDS from \citet{will09} covering $BVRIz$
from Subaru/SuprimeCam, $JK$ from UKIDSS, and deep \spitzer/IRAC
imaging from the \spitzer\ UDS covering 3.6-8.0~\micron.  Our catalog
is identical to that used in \citet{papo10a}.    In summary, these catalogs reach $5\sigma$
limiting magnitudes of $B < 27.7$, $R < 27.1$, $i < 26.8$, $z < 25.5$,
$J< 23.9$ and $K < 23.6$ mag in apertures of 1\farcs75 diameter.  The
catalogs are selected in the $K$-band with colors measured in fixed
apertures and they include unique aperture corrections for each source
defined as the difference between the fixed aperture and total
magnitude derived from the K-band for that source
\citep[see][]{will09}.   We used IRAC catalogs with photometry
measured in 4\arcsec\ diameter apertures, with aperture corrections to
total magnitudes as stated in \citet{papo10a}.  We matched the IRAC
sources to the optical+IR data within 1\arcsec\ radii.   The final catalog
covers a wavelength baseline of 0.4--8~\micron.

We use the photometric redshift catalog from
\citet{papo10a} derived using EAZY v1.0 \citet{bram08}.   In summary,
we used the default galaxy spectral energy distribution templates with
the default $K$-band prior based on the luminosity functions of
galaxies in a semi-analytic simluation.  In addition to photometric
redshifts, we measure the full photometric-redshift probability
distribution function, $P(z)$, normalized such that $\int\, P(z)\, dz
= 1$ when integrated over all redshifts.  As discussed in
\citet{papo10a}, compared to existing spectroscopic redshifts, the
best-fit photometric redshifts have uncertainties of $\Delta([z_{sp} -
z_{ph}]/[1 + z_{sp}]) = 0.04$ derived from the normalized median
absolute deviation \citep{beers90} in the range $1 \leq z_{sp} \leq
2$ (which includes the redshift-range of interest in this paper).

We use stellar masses, star-formation rates (SFRs)
and specific SFRs (sSFR, the SFR per unit stellar mass), and
rest-frame $U-B$ and $U-V$ colors as derived in \citet{papo12}.   To
summarize, we fit the 10-band galaxy photometry with model spectral
energy distributions with Chabrier initial mass function and solar
metallicity using stellar population synthesis models from
\citet{bruz03}, allowing for a range of stellar population age,
star-formation history and dust attenuation using the \citet{calz00}
law.   The rest-frame $U-B$ and $U-V$ colors and SFRs are derived
using the best-fit models, where we measure the SFR averaged over the
prior 100 Myr.   We found that the rest-frame colors for galaxies in
our sample agree with those
independently derived from the best-fitting templates for the
photometric redshifts using EAZY with a median rest-frame color
differences of $\Delta(U-B) = 0.03$ and $\Delta(U-V)=0.04$ mag with
standard deviations  $\sigma(\Delta[U-B]) = 0.04$ and
$\sigma(\Delta[U-V])=0.08$ mag.  

We adopt for our galaxy sample the limiting magnitude from the
\citet{will09} catalog of $K = 22.9$~mag. This equates approximately to
a 10$\sigma$ detection limit.  This magnitude limit for a galaxy at
$z=1.6$ corresponds to a stellar mass of $\approx 2\times
10^{10}$~\msol, for a solar metallicity stellar population with a
maximum mass-to-light ratio (formed in a burst at $z_f = 5$). Galaxies
with younger stellar populations will have lower mass-to-light ratios.
Therefore, while our sample is ``complete'' for all galaxies with
stellar masses above $\approx 2\times 10^{10}$~\msol, our sample is
sensitive to galaxies with younger stellar populations with stellar
masses below this limit.  Therefore, our stellar mass limit is
``conservative'' in the sense that we are sensitive to all galaxies
above this mass limit.  The choice for this limiting stellar mass is
also motivated by our ability to recover accurate morphological
information from the \hst/CANDELS images with GALFIT.  Galaxies 
at the stellar mass limit and with the fiducial stellar population above have
magnitudes of  $m(\mathrm{F125W}) = 24$~mag, up to which we can recover robust
quantitative morphological parameters with accuracies better than 40\%
(see below, \S~\ref{section:analysis:sersic} and the Appendix).   We
therefore adopt a limiting mass of $2\times 10^{10}$~\msol\ where we
are complete for all galaxies, and for which our analysis is robust.

We focus on the subset of the UDS with CANDELS imaging.    In addition
to the portion of the UDS covered by CANDELS, we find it is useful to
include the full UDS dataset when making comparisons not involving
size or morphology. This larger data set covers an area of 0.70
deg$^2$ (as discussed in \citet{will09} and
\citet{papo12}) and is nearly $10\times$ larger in area than the CANDELS field.      

As in \citet{papo12} we select galaxies around the redshift $z=1.6$
by defining the integrated photometric redshift probability
distribution function, $\mathcal{P}_z \equiv \int\,\, P(z)\, dz$
integrated of the redshift range $z = z_\mathrm{cen} \pm \delta z$
with $z_\mathrm{cen} = 1.625$ \citep[which corresponds to the mean
redshift of the cluster in this field]{papo10a} and $\delta z = 0.05
\times (1+z_\mathrm{cen})$, which is typical of the photometric
redshift 68\% confidence limits at this redshift. Here we select all
galaxies with $\mathcal{P}_z=0.4$. As shown in
\citet{papo10a}, this $\mathcal{P}_z$ limit is required to include
blue, star-forming galaxies with spectroscopically confirmed redshifts
at $z\sim 1.6$, and it includes the red, quiescent galaxies at this
redshift.  Using a  higher $\mathcal{P}_z$ limit restricts the sample
more to quiescent galaxies (as was used by Papovich et al.\ 2012).
Selecting galaxies this way has the advantage that it uses all
information about the galaxies' photometric redshift probability
distribution functions.   In contrast, using samples of galaxies
selected solely with a best-fit photometric redshift excludes
information, and galaxies selected in some redshift interval will be
very dependent on the errors in the photometric redshifts and subject
to biases.   Regardless, in the case here we compare the properties of
both cluster and field galaxy samples constructed from the same
photometric-redshift selection and using the same CANDELS data
set. Therefore, any systematics or biases resulting from the selection
affect both samples equally.   As a result, the relative comparison
between the galaxies in the cluster and field is robust. 

We select all galaxies with $\mathcal{P}_z > 0.4$.   This selection
provides a sample of 433 galaxies with coverage in CANDELS imaging
with mean (best-fit photometric) redshift $\bar{z} = 1.61$ with a
standard deviation of 0.12. 

\section{Analysis}

\subsection{Quantifying Galaxy Morphologies}\label{section:analysis:sersic}
 
We used GALFIT \citep{peng02} to fit  parametric models to the surface
brightness profiles of our sample galaxies in the WFC3 F125W imaging.
We chose the F125W bandpass for this analysis because it  corresponds
approximately to the $B$-band in the rest-frame at
$z=1.6$. Many studies of the morphologies of  distant
galaxies (especially in clusters) are based in the rest-frame $B$-band
\citep[see, e.g.,][]{blak03,homei05,post05,holden05,blak06,mei09,santos09}
, and our choice of the F125W bandpass here allows for direct
comparisons with these other studies.  In addition, the CANDELS
imaging allows for  studies of galaxy morphology in the rest-frame
$B$-band out to $z \sim 3$ using the F160W bandpass, which would
permit the extension of our work to higher redshift. Therefore our
choice here seems prudent.   Regardless, our tests (see below) show
that none of our conclusions would be changed if we used the F160W
bandpass instead.     

GALFIT models the surface brightness profile of a galaxy as
$\exp(-R/R_\mathrm{eff})^{1/n}$ \citep{sersic68}, where $n=1$
corresponds to a exponential (disk) profile and $n=4$ corresponds to the
de Vaucouleur's profile.  GALFIT convolves these models with
a user-defined PSF. We used a PSF constructed using TinyTim v7.2
\citep{krist95} and combined with the same dither positions as the
CANDELS data.   For each galaxy we fit with GALFIT the  effective semimajor axis
($a_\mathrm{eff}$), background, axis ratio ($b/a$), total magnitude,
and \sersic\ indices ($n$)  as free parameters.  

We compared our \sersic\ index  measures using the F125W imaging for
the galaxies in our $z\sim 1.6$ sample against independent
measurements derived using GALAPAGOS \citep{barden12} with the CANDELS
WFC3/F160W imaging as described by  \citet{bell12} and \citet{vanderwel12}.
We find a median offset in \sersic\ index of $-0.03$
dex in $\Delta(n)/n$ with a scatter of 0.12 dex and we find a median
offset in effective radius of 0.01 in $\Delta(r_\textrm{eff}) /
r_\textrm{eff}$ with a scatter of 0.06 dex.     The uncertainties are
nearly independent of galaxy magnitude, with only minor improvement for $J <
23$~mag compared to $23 < J < 24$~mag.   Because these differences
are small, our conclusions would be unchanged if we used fits from the
F160W imaging instead of the F125W imaging.

We performed a series of simulations, inserting model galaxies
into the \hst\ data, and using GALFIT as described above to recover
their  parameters. Our analysis of these simulations showed that the
recovered effective radii are accurate to better than 40\% and the
\Sersic\ indices to better than 25\% for galaxies with $\reff=2$ kpc
and $n=4$ with magnitude $m(F125W)=24$ mag,  near the stellar-mass
limit of our sample, ($2 \times 10^{10}$~\msol, see \S~2). Galaxies
with parameters such as these represent the worst case scenario, and
errors for most objects will be significantly smaller than this.
The uncertainties are correlated between parameters, similar to the findings of
\citet{haus07}. For more information, see the appendix.

\ifsubmode
\begin{figure}
\epsscale{1.0}
\else
 \begin{figure*}
\epsscale{1.0}
\fi
\plotone{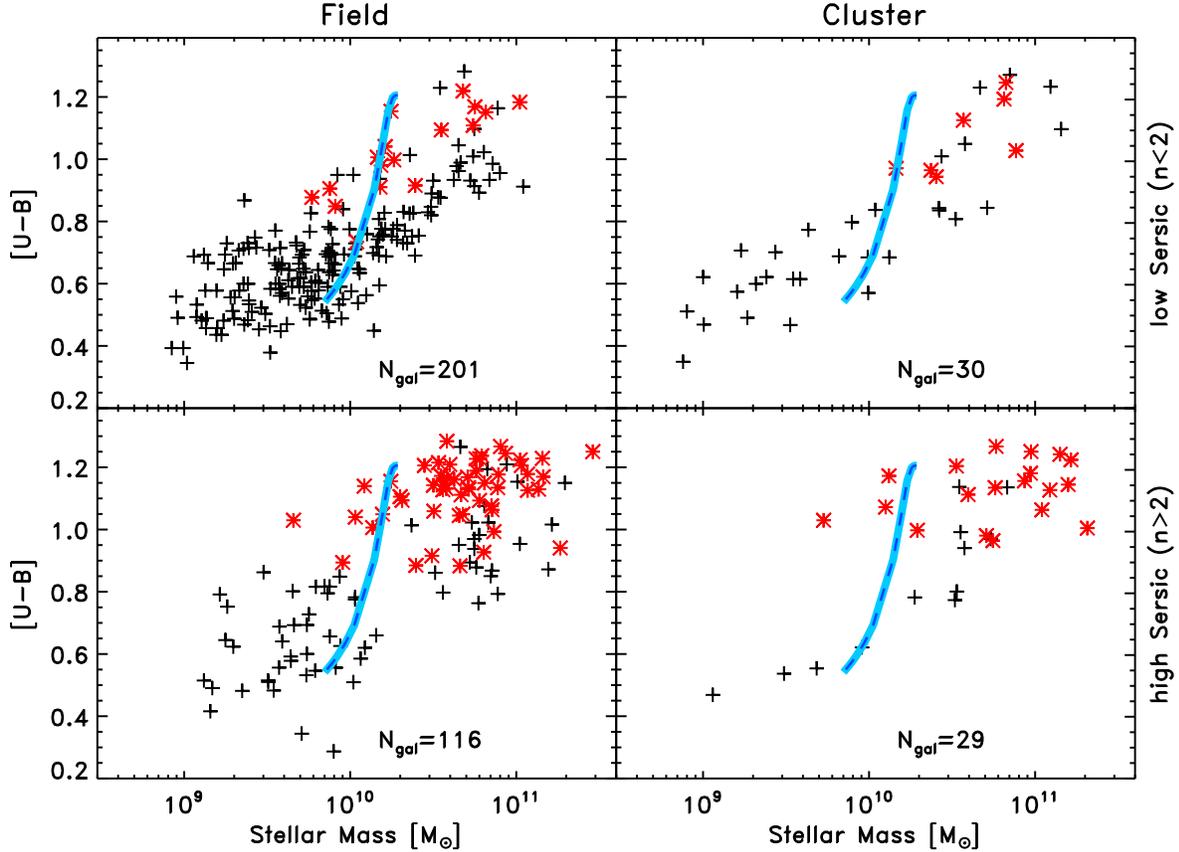}
\epsscale{1.0}
\vspace{12pt}
 \caption{The rest-frame $U-B$ color -- stellar mass relation for
galaxies at $z=1.6$.    Each panel shows a subset of galaxies in the
field and cluster with high and low \sersic\ index.   The left column
shows the relation for galaxies in the field, $\rproj > 3$~Mpc from
the cluster center.  The right column shows the relation for the
cluster, with $\rproj < 1.5$~Mpc.   The upper row shows the relation
for the subsample with low\ sersic\ indices ($n < 2$), and the lower
row shows the subsamples with high \sersic\ indices ($n > 2$).
Red-colored points denote objects that satisfy the color selection for
quiescent galaxies.  The thick blue lines denote the estimated
stellar--mass completeness for stellar population with a maximal
mass-to-light ratio, which formed in burst at $z_f=5$ with solar
metallicity and subsequent passive evolution.  Our analysis is
complete for all galaxies with stellar masses $>2 \times
10^{10}$~\msol, where our analysis shows our
quantitative morphological parameters are
robust. } \label{fig:colormass}
\ifsubmode
\end{figure}
\else
\end{figure*}
\fi
  
\subsection{Quantifying Galaxy Environment}\label{section:analysis:environment}

We tested different ways to characterize the environment of each
galaxy in our $z=1.6$ sample.   We found that using samples based on
the projected distance from the cluster center provided the most
significant results.   For this reason, we defined two samples.  We
selected a sample of ``cluster'' galaxies with $\mathcal{P}_z > 0.4$
and projected distances $\rproj < 1.5$~Mpc of the cluster center
\citep[with the cluster center defined by][]{papo10a}. A subsample of 
``field'' galaxies with $\rproj > 3$~Mpc was selected for comparison.

We also calculated environment measures based on the projected distance to the
$N$th nearest neighbor, $D_N$ \citep[\eg,][]{dres80}.  We computed the
$N$th nearest neighbor distance from each galaxy in our $\mathcal{P}_z
> 0.4$, $z\sim 1.6$ sample to (1) each other galaxy
in the $z \sim 1.6$, $\mathcal{P}_z > 0.4$ sample, as well as to (2)
each galaxy in a sample that have \textit{best-fit} photometric
redshifts $1.50 < z < 1.75$.  Additionally, we tested the $N$th nearest
neighbor distance with the requirement that the stellar-mass ratio between
each galaxy and its neighbors be greater than or equal to a certain
value over a range from 1:1 to 1:10
\citep[as proposed by][]{haas11}.  In practice, we
find that using a nearest-neighbor distance estimator, e.g., the
$N=2$nd-nearest neighbor with $D_2 < 0.25$~Mpc, identified the
majority of galaxies with $\mathcal{P}_z > 0.4$ within $\rproj \lsim
1$~Mpc of the cluster.  However, \textit{all} of the $D_N$ estimators
selected only a small fraction ($\lsim 20$\%) of quiescent galaxies
with $\mathcal{P}_z > 0.4$ within 1 Mpc $< R <$1.5 Mpc from the
cluster:  the $N$th nearest neighbor distances of these galaxies are
simply too large to place them in high local number
density regions based on this statistic. 
\citet{haas11} showed that including a requirement on the stellar-mass
ratio between neighboring galaxies in the density estimator improves
the strength of the correlation between halo mass and density.
However, our tests showed that this offers no improvement.  Indeed, by
requiring neighboring galaxies to be at least as massive as the galaxy
in question, nearly all of the most massive galaxies in the
cluster are shifted to ``low'' density regions because there are
\textit{no} other galaxies of comparable mass in the sample (and
therefore the distance to the $N$th nearest neighbor with comparable
mass becomes very large). 

It has been shown that $N$th nearest neighbor measurements are good
probes of internal cluster properties and galaxy-group-sized halos
while fixed aperture methods better examine the cluster as a whole
\citep{muldrew12}.  For this reason, \citet{woo12} conclude that using
the relative projected distance of a  satellite from its halo center
is better suited to studying processes associated with
clustering on the largest scales than $N$th-nearest neighbor
statistics.  Moreover, we are motivated by the results from other
studies that have found that the environment affects galaxies as they
become satellites of larger dark-matter halos, whereas the evolution
of central galaxies depends primarily on the halo mass (see \S 1).
Therefore, here we adopt the projected distance from the
cluster center as our measure of environment for our study.  We define our
samples as the ``cluster'' (with $\rproj < 1.5$~Mpc) and ``field''
(with $\rproj > 3$~Mpc). In addition, we will divide the ``cluster''
sample spatially into two subsamples: the cluster center with
$R_\mathrm{proj} < 1.0$ Mpc and the cluster outskirts defined as the
annulus with 1.0 Mpc $< R_\mathrm{proj} <$ 1.5 Mpc.

\section{THE RELATION BETWEEN ENVIRONMENT, MORPHOLOGY, COLOR AND MASS AT $z=1.6$}\label{section:cccut}

Figure \ref{fig:colormass} shows the relation between the $U-B$
rest-frame color and stellar mass for galaxies in our sample in the
cluster ($\rproj < 1.5$~Mpc) and the field ($\rproj > 3$ Mpc), divided
into  subsamples of disk-dominated galaxies (\sersic\ index  $n < 2$)
and bulge-dominated galaxies (\sersic\ index $n > 2$).

In all panels a strong $U-B$ color--mass, and color--mass-morphology
relation is apparent.   Galaxies with high stellar mass have red
colors \citep[similar to,][]{will09,quadri12}.    The \Sersic\
index is correlated with mass and color.   Exactly half ($82/164$) of
bulge-dominated ($n >2$) galaxies in our sample lie on the
red sequence, and the majority ($238/269=88.5$\%) of disk-dominated
galaxies ($n < 2$) reside in the blue cloud.
Disk-dominated galaxies populate the full range of $U-B$ color.  This
is similar to the findings of previous studies at this redshift, for
example, \citet{wuyts11} and \citet{bell12}, and references therein. 

Figure~\ref{fig:colormass} also shows the color-mass-morphology
distributions for galaxies with colors consistent with older, passively
evolving stellar populations.   We selected these ``quiescent'' galaxies
using a variant of the $UVJ$ selection \citep{wuyts07,will09} based on
the observed $z-J$ and $J-\mone$ colors as defined by \citet{papo12}, 
\begin{eqnarray}\label{eqn:colorselection}
(z - J)_\mathrm{AB}\bs&\ge&\bs 1.3\,\,\mathrm{mag}\nonumber \\
(J - \mone)_\mathrm{AB}\bs& \le&\bs 2.1\,\,\mathrm{mag} \\
(z - J)_\mathrm{AB}\bs&\ge&\bs 0.5 + 0.55 (J - \mone)_\mathrm{AB}. \nonumber
\end{eqnarray} 

This color-selection classification is efficient in isolating galaxies
with very different specific SFRs. In \citet{papo12} we showed that
using this color selection, the majority ($\sim$90) of quiescent
galaxies selected have specific SFRs $<10^{-2}$Gyr$^{-1}$. Therefore,
while their SFR may not be zero, it is highly ``suppressed''
\citep[\eg,][]{kriek06}. In contrast, galaxies that do not
satisfy our definition for ``quiescent'' have specific SFRs $\sim$ 1
Gyr$^{-1}$, at least one order of magnitude higher, which are
characteristic of galaxies on the star-forming ``sequence''
\citep[\eg,][]{daddi07a,noeske07a}. Therefore, although strictly
speaking the color-selection criteria in
equation~\ref{eqn:colorselection} separates galaxies exhibiting highly suppressed
star-formation from those that do not. For brevity, we refer to these
two samples as ``quiescent'' and ``star-forming'' galaxies respectively.

For much of the analysis here, we use a one-sided
Wilcoxon-Mann-Whitney rank-sum (WMW) test \citep{mann47} to place a
significance on the likelihood that the cluster and field galaxies
have similar properties and are drawn from the same parent
population.  A summary of these likelihoods is given as
probabilities  ($p$-values) in Table~\ref{table}. 

\subsection{Do the $U-B$ Colors Depend on Environment?} \label{section:umb}

 We do observe a slight increase in the $U-B$ color of galaxies
associated with the cluster compared to the field.  This is
illustrated in Figure~\ref{fig:umb}, which shows the cumulative
distribution of $U-B$ color for both samples using the data for the
CANDELS subset.   The WMW test gives that the CANDELS galaxies in the
cluster have higher $U-B$ color, with a WMW significance $p = 0.069$.
Using the larger full UDS sample improves this significance, with $p =
0.033$ ($\simeq 2\sigma$).    

Most of the difference in the $U-B$ colors stems from differences in
the colors of the star-forming galaxies.   The WMW test for the $U-B$
colors finds no difference in the distributions for  the quiescent
galaxies ($p \gsim 0.3$ for all samples).    The star-forming galaxies
in contrast show slight evidence that the $U-B$ colors of the galaxies
in the cluster are redder, although the significance is weak with $p =
0.07$ (about $1.5\sigma$).  There is little difference in this
$p$-value for the comparison between the smaller CANDELS samples and
the larger UDS samples.  

If this difference in the
$U-B$ colors of the star-forming galaxies is real, then it would
provide evidence that the star-forming galaxies associated with the
cluster have more dust extinction, which reddens the colors.
This could imply that the SFRs we derive from the SED fitting may be lower
limits, missing highly extincted star-formation, which is indicated by
the redder $U-B$ colors. 

For reasons discussed below, we tested if galaxies at larger clustercentric distances, defined by
the annulus 1 Mpc $< \rproj <$1.5 Mpc,  have different 
$U-B$ colors compared to those galaxies within
$\rproj < 1$~Mpc from the cluster.  However, the WMW test gives no measureable
difference with $p$-values $> 0.11$ for all
subsamples, which is not significant. Therefore, the cluster galaxies
within $\rproj < 1$~Mpc have similar $U-B$ colors as those galaxies
within 1 Mpc $<\rproj < $1.5~Mpc.      We repeated this test for only
the subsample of galaxies with $M=(2-9) \times 10^{10}$ \msol, but find no substantial
differences.

\subsection{Do the specific SFRs of
  Galaxies Depend on Environment?} \label{section:sfrs}

Considering the combined samples (quiescent and star-forming
galaxies), the WMW does not rule out the hypothesis that the specific
SFR distributions are drawn from the same parent populations (all have $p >
0.22$), and we see no difference in the specific SFR distributions for
more restrictive samples of quiescent or star-forming galaxies.    We also
see no significant difference using either the smaller CANDELS sample
or larger UDS sample.   There is no apparent difference in the
specific SFR distributions between the cluster and the field, nor is
there any significant difference in the distributions for galaxies in
the cluster core ($R_\mathrm{proj} < 1$ Mpc) and those in the
$R_\mathrm{proj} = 1-1.5$ Mpc annulus.

On the surface this result appears to contrast some studies that find
an increase in the SFR (or SFR surface density) of galaxies in the
cores of clusters at this redshift \citep[\eg,][]{tran10,koce11}. For
example, \citet{tran10} found an elevated IR luminosity projected
surface density associated with this cluster compared to the field. We
attribute this apparent discrepency to the fact that in our study we
have not included IR data, and therefore our SFRs (and sSFRs) may be
underestimated. Indeed, this is consistent with the
star-forming galaxies in the cluster having redder U-B rest frame colors
from \S 4.1. As discussed above, this may imply these galaxies have
high extinction, which would imply higher extinction-corrected SFRs
that we are missing in our measurements. 

\subsection{Do the Stellar Masses Depend on Environment?} \label{section:mass}

In general, there is no substantial difference in the distribution of
stellar masses when all galaxies (quiescent and star-forming) are
considered in the field and the cluster.   For the more restrictive
sample of quiescent galaxies only, those  associated with the cluster
show a possible increase in stellar mass relative to the field. The
WMW test gives a significance $p=0.065$ for the CANDELS samples.
However, this increases to $p=0.078$ when all galaxies in the larger
UDS samples are considered, implying a weaker significance.    Therefore, there
is slight evidence ($1.5\sigma$ significance) that quiescent
galaxies associated with the cluster have larger stellar masses
compared to those in the field. The difference appears driven
by the higher stellar masses of the quiescent galaxies in the cluster
compared to the field, which is evident in Figure \ref{fig:colormass}.

\ifsubmode
\begin{deluxetable}{lcccccccc}
\else
\begin{deluxetable*}{lccccccc}
\fi
\tablecaption{Summary of Wilcoxon-Mann-Whitney $U$
  probabilities\\comparing the properties of 
  $z=1.6$ cluster and field galaxies in the CANDELS UDS field\label{table}}
\tablecolumns{8}
\tablewidth{6in}
\tablehead{
\colhead{Sample} & 
\colhead{$N_\mathrm{cluster}$} & 
\colhead{$N_\mathrm{field}$} & 
\colhead{$p(U-B)$} & 
\colhead{$p(\log M_\ast)$} & 
\colhead{$p(\log$ sSFR)} & 
\colhead{$p(\reff)$} & 
\colhead{$p(n)$} \\
\colhead{(1)} & 
\colhead{(2)} & 
\colhead{(3)} & 
\colhead{(4)} & 
\colhead{(5)} & 
\colhead{(6)} & 
\colhead{(7)} & 
\colhead{(8)}
}

\startdata
\multicolumn{1}{l}{All Galaxies} & 
\multicolumn{7}{l}{$\log M_\ast / \msol > 10.30$} \\ [2pt]
UDS & 56 & 2061 & 0.0331 & 0.125 & 0.296 & \ldots & \ldots \\
CANDELS & 37 & 110 & 0.0692 & 0.177  & 0.218 &  0.422 & 0.0800 \\ [5pt]

\multicolumn{1}{l}{All Galaxies} & 
\multicolumn{7}{l}{$10.30 < \log M_\ast / \msol < 10.95$} \\ [2pt]
UDS & 43 & 1701 & 0.100  & 0.308 & 0.412 & \ldots & \ldots \\
CANDELS & 27 & 94 & 0.229  & 0.390 & 0.263 & 0.493 & 0.091 \\ [5pt] \hline
 
\multicolumn{1}{l}{Quiescent Galaxies} & 
\multicolumn{7}{l}{$\log M_\ast / \msol > 10.30$} \\ [2pt]
UDS & 35 & 1083 & 0.458 & 0.0781 & 0.216 & \ldots & \ldots \\
CANDELS & 21 & 51 & 0.397 & 0.0646 & 0.202 & 0.0571 & 0.0234 \\ [5pt]

\multicolumn{1}{l}{Quiescent Galaxies} & 
\multicolumn{7}{l}{$10.30 < \log M_\ast / \msol < 10.95$} \\[2pt]
UDS & 24 & 870 & 0.291 & 0.365 & 0.292 & \ldots & \ldots\\ 
CANDELS & 13 & 41 & 0.292 & 0.310 & 0.420 & 0.107 & 0.0164 \\ [5pt] \hline

\multicolumn{1}{l}{star-forming galaxies }  & 
\multicolumn{7}{l}{$\log M_\ast / \msol > 10.30$} \\ [2pt]
UDS & 21 & 978 & 0.0699 & 0.466 & 0.409 & \ldots & \ldots \\
CANDELS & 16 & 59 & 0.0737 & 0.339 & 0.193 & 0.280 & 0.238  \\ [5pt]

\multicolumn{1}{l}{star-forming galaxies }  & 
 \multicolumn{7}{l}{$10.30 < \log M_\ast / \msol < 10.95$} \\ [2pt]
UDS & 19 & 831 & 0.0586 & 0.353 & 0.444 & \ldots & \ldots \\
CANDELS & 14 & 53 & 0.137 & 0.254 & 0.297 & 0.181 & 0.457

\enddata \tablecomments{(1) Description of samples used,  (2) number
of cluster galaxies in WMW test, (3) number of field galaxies used in
WMW test. Other columns are the WMW significance of the difference in
the cluster and field samples between
the distributions of (4) stellar-mass, (5) specific SFR, (6) effective radii, (7)
\sersic\ indices.  Low probabilities ($p$-value)
indicate more significant differences in the distributions. } \ifsubmode
\end{deluxetable}
\else
\end{deluxetable*}
\fi

\ifsubmode
\begin{figure}
\epsscale{1.0}
\else
\begin{figure}[t]
\epsscale{1.15}
\fi
\plotone{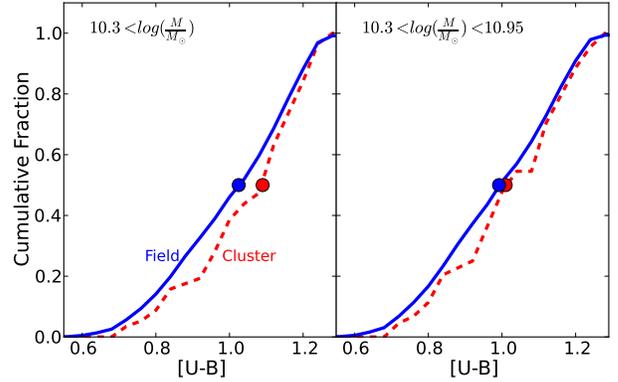}
\ifsubmode
\else
\fi
\epsscale{0.75}
\caption{The cumulative distribution of rest-frame $U-B$ colors for
all galaxies in the field ($\rproj > 3$~Mpc, solid blue line) and in
the cluster ($\rproj < 1.5$~Mpc, dashed red line). Each curve is
normalized to the total number of galaxies in that sample (these
curves include all galaxies in the UDS, not only those in the CANDELS
field).  The left panel shows the distribution for all quiescent
galaxies with stellar masses $>2 \times 10^{10}$~\msol.   The circle
points on each curve show the median value of the distributions.
There is weak evidence that the galaxies associated with the 
cluster have a median $U-B$ color that is $\Delta(U-B) \sim 0.1$ mag
redder than those in the field.   The WMW gives $p = 0.033$ for the
total UDS samples, but declines to $p=0.069$ using only the CANDELS
samples.   The right panel shows a subset with moderate stellar mass
$2\times 10^{10} - 9 \times 10^{10}$~\msol.  The WMW test finds a
reduced significance, $p > 0.100$  (see Table~\ref{table}).\\ }\label{fig:umb}

\ifsubmode
\end{figure}
\else
\end{figure}
\fi

The stellar masses of galaxies in the 1 Mpc $ < \rproj <$ 1.5 Mpc
annulus show evidence of being different from those
with $\rproj < 1$~Mpc.  Considering all galaxies (star-forming and
quiescent), the galaxies within $\rproj < $1 Mpc have stellar masses
approximately a factor of 1.9 higher than those within 1 Mpc $< \rproj
<$ 1.5 Mpc (with WMW $p = 0.0053$). 
This significance is reduced ($p = 0.078$) when considering only
quiescent galaxies.
Interestingly, there is no difference between the galaxies within 1
Mpc $< \rproj <$ 1.5 Mpc and those in the field, $\rproj > 3$ Mpc, ($p
> 0.2$), suggesting galaxies in this annulus share a mass distribution
more characteristic of the field at this redshift than the
highest-density region associated with the cluster.
This is possible evidence that these galaxies were
recently members of the field and are in the process of being
accreted into the cluster environment.

\subsection{Do the \sersic\ Indices of Galaxies\\Depend on
Environment?}\label{section:sersic}

We now compare the difference in \sersic\ indices between the cluster
and field populations in more detail. This is illustrated
in Figures \ref{fig:radec} and \ref{fig:serrad}.
Figure \ref{fig:radec} shows the astrometric positions of quiescent
and star-forming galaxies at $z=1.6$ in the CANDELS field.  The
symbols correspond to the \sersic\ index of the quiescent galaxies.
Figure \ref{fig:serrad} shows the \sersic\ indices of galaxies in the
CANDELS field as a function of their clustercentric distance.
These figures show that there appears to be an excess of galaxies with
$n < 2.5$ in the annulus 1 Mpc $< \rproj < $1.5 Mpc from the cluster
relative to either the sample with $\rproj <1$ Mpc or the field sample
($\rproj >$3 Mpc), and that this effect is driven by intermediate mass
galaxies. It is also apparent that quiescent galaxies with $n<1$ are
only found in the 1 Mpc $< \rproj <$1.5 Mpc annulus.   As discussed in \S~3.1, to test if this is a
result of our choice of F125W bandpass or analysis, we checked our
\sersic\ indices against those computed independently from the F160W
data from \citet{bell12}, and we found no change. 

\subsubsection{Comparison Between the Cluster and the Field}

Table~\ref{table} shows the WMW $p$-values comparing the properties
of galaxies in the cluster and field environments.    Based on
our analysis there is no evidence that star-forming galaxies in the
field and cluster have differences in their \sersic\ indices.   

We do observe differences in the distributions of \sersic\ indices for quiescent galaxies. 
There is an envelope in the \sersic\ index and stellar mass
distribution for quiescent galaxies:  more massive quiescent galaxies
generally  have higher \sersic\ indices, and only a few of the most massive
($> 9 \times 10^{10}$~\msol) galaxies have $n < 2$
(Figure~\ref{fig:colormass}). 
Most, 81/108 = 75\%, of the quiescent galaxies in both the field and
cluster have high \sersic\ indices, $n > 2$, and are
bulge-dominated \citep[consistent with][]{bell12,wuyts11,papo12}.
However, more notably, the fraction of galaxies with \textit{low}
\sersic\ index $n < 2$ that are quiescent is significantly higher in
the cluster (6/16=38\%) compared to the field (8/43 = 15\% ), for
galaxies 
above our mass limit. This is
significant at about $2.5\sigma$ from binomial statistics, although it
is based on small numbers of galaxies.  As illustrated in Figures
\ref{fig:radec} and \ref{fig:serrad} and as we discuss below, most of
this difference comes from galaxies in the annulus 1 Mpc $< \rproj <$
1.5 Mpc from the cluster, and suggests some mechanism suppresses
star-formation in galaxies with low \sersic\ index at large
clustercentric radii.

\ifsubmode
\begin{figure}
\epsscale{1.0}
\else
\begin{figure*}[t]
\epsscale{1.}
\fi
\plotone{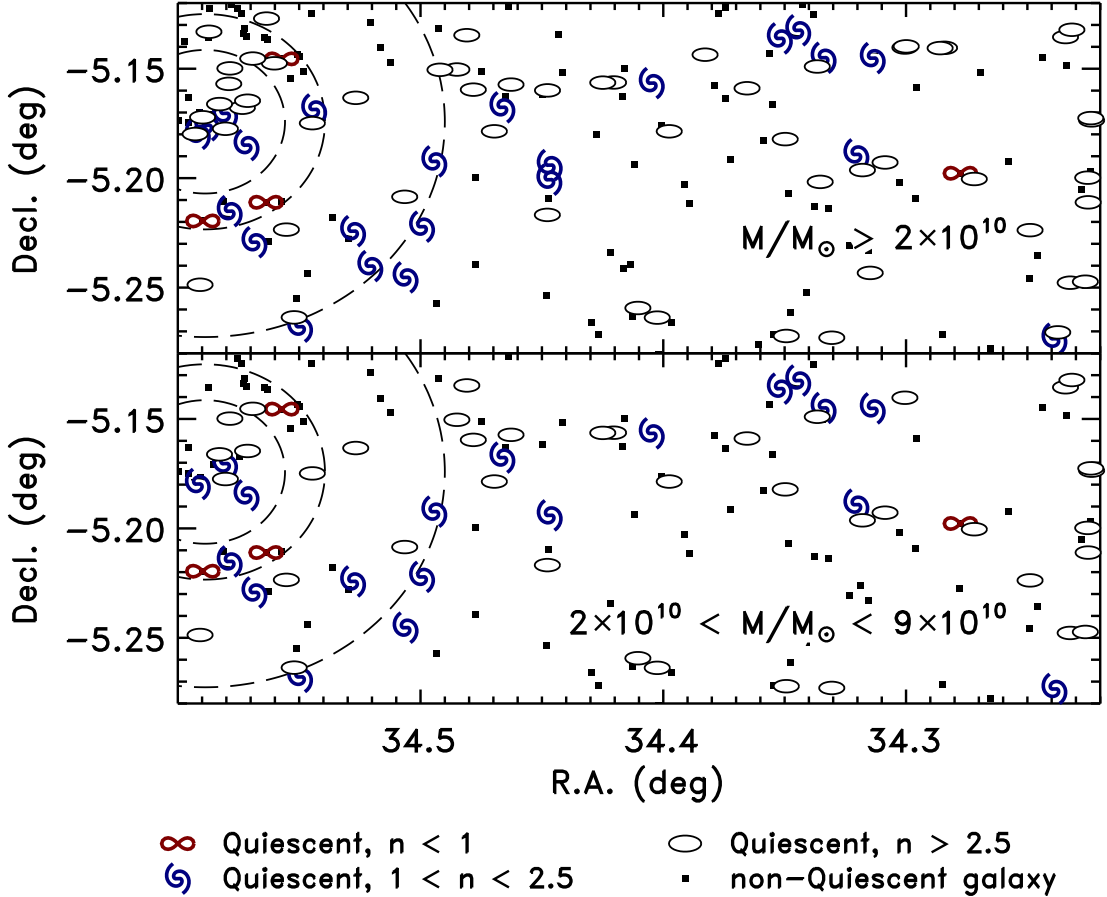}
\ifsubmode
\else
\fi
\epsscale{1.0}
\caption{The spatial distribution of galaxies at $z=1.6$ in the
  CANDELS UDS field.   Symbols denote quiescent (large symbols) and
  star-forming galaxies (small boxes), as labeled.    For the
  quiescent galaxies, the symbol type corresponds to \sersic\ index,
  as indicated in plot legend.    The concentric dashed circles show
  projected distances of $\rproj = $1, 1.5, and 3 Mpc from the
  astrometric center of the cluster using the coordinates of
  \citet{papo10a}.    The quiescent galaxies in the annulus 1 Mpc $<
  \rproj < 1.5$~Mpc have lower \sersic\ indices and larger effective
  sizes compared to other quiescent galaxies.  } \label{fig:radec}
\ifsubmode
\end{figure}
\else
\end{figure*}
\fi 

The WMW probabilities in Table~\ref{table} show the evidence for
differences in the \sersic\ indices of galaxies associated with the
cluster and the field.   Most of this difference is
isolated to the quiescent galaxies:  star-forming galaxies in both the
field and cluster have median \sersic\ indices $n_\mathrm{med} \simeq
1.4$, with no significant differences, $p > 0.2$ .
Quiescent galaxies associated with the cluster have lower median
\sersic\ indices, $n_\mathrm{med}=2.6$ versus
$n_\mathrm{med}=3.3$ in the field, and the WMW test shows this is
significant with $p = 0.023$ (approx.\ $2\sigma$)
considering the full sample with $M > 2 \times 10^{10}$~\msol.
Figure~\ref{fig:cumsersic} shows the cumulative distribution of the
\sersic\ indices of quiescent galaxies in the field and the cluster.

The right panel of Figure 3 shows the cumulative distributions for
the more restrictive subsamples of quiescent galaxies with moderate
stellar mass, $2\times 10^{10} - 9 \times 10^{10}$~\msol.   Here the median
\sersic\ index is $n_\mathrm{med} = 2.4$ for the cluster galaxies, and
$n_\mathrm{med} = 2.8$ for the field galaxies, and the significance
increases to $p = 0.016$. 

This result has a significance of $\approx2\sigma$, and is based on relatively small
samples:  21 quiescent galaxies with stellar mass $> 2\times
10^{10}$~\msol\ which lie within $\rproj < 1.5$~Mpc from the cluster
as covered by CANDELS (table~\ref{table}).   To test further the
significance of this result, we used a bootstrap Monte Carlo test to
estimate a likelihood that we obtained the observed difference in
\sersic\ indices by chance.  We used as a parent sample the combined
samples of all quiescent galaxies in the cluster and field.  We then
constructed 10,000 pairs of cluster-sized and field-sized samples of
$N$(cluster) and $N$(field) objects randomly selecting from the parent
sample (with replacement), where $N$(cluster) and $N$(field) are equal
to the size of the original cluster and field samples, respectively.
We then recomputed the WMW $p$ value between each pair of random
cluster and field samples.  In this test, we obtained a $p$ value as
small (or smaller) than the one we observe in only 0.6\% of the random
samples when drawing from the full sample of galaxies with $M > 2
\times 10^{10}$~\msol\  (this increases slightly to 0.9\% of cases for
$M = 2 \times 10^{10} - 9\times 10^{10}$~\msol\ moderate-mass
samples). Therefore, the likelihood that we would have obtained our
result by chance if there were no actual difference between the
cluster and field objects is smaller than 1\% (equivalent to the
probability of $>2.6\sigma$ significance for a Gaussian
distribution).

\ifsubmode
\begin{figure}
\epsscale{1.0}
\else
\begin{figure*}[t]
\epsscale{1.}
\fi
\plotone{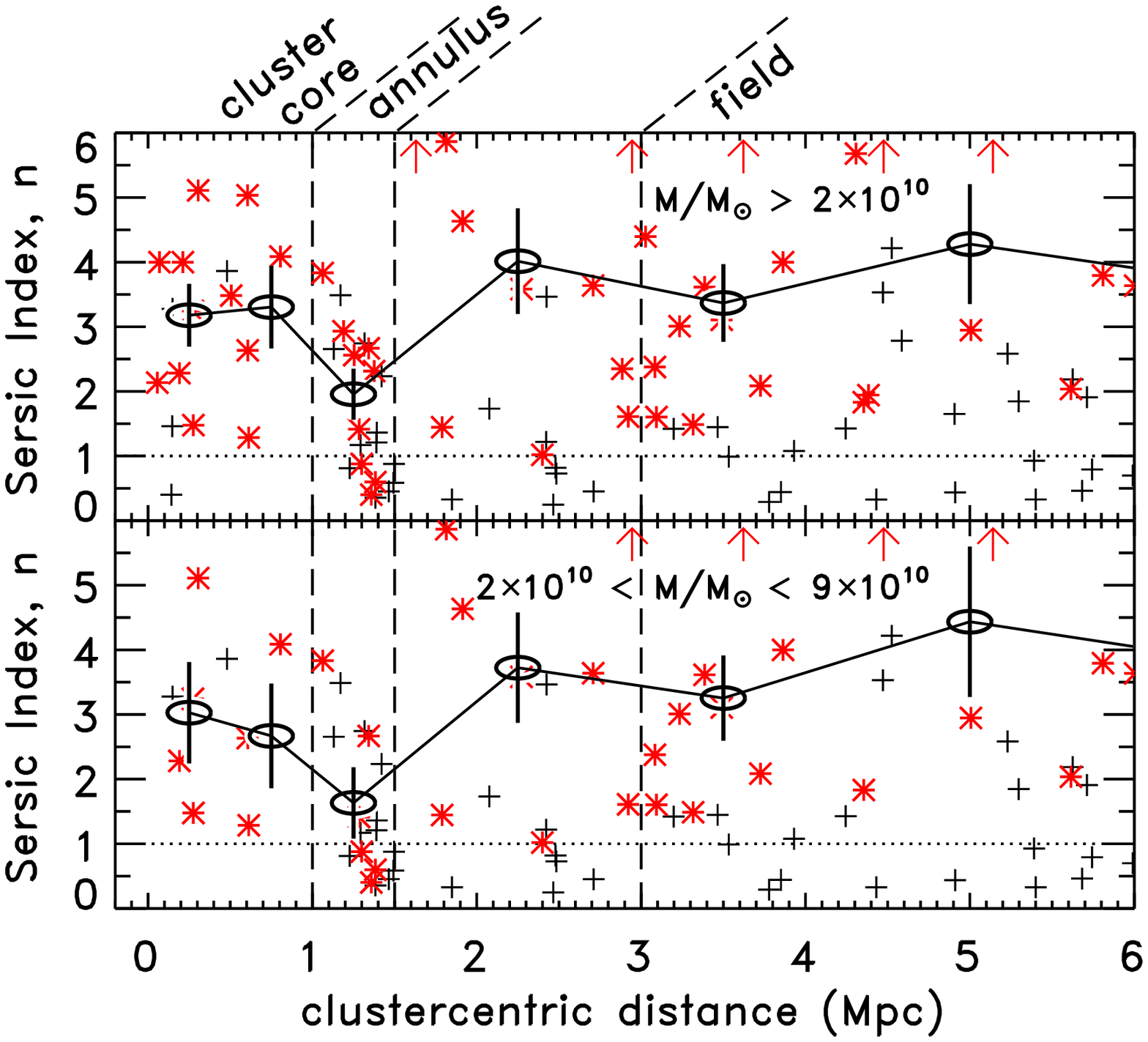}
\ifsubmode
\else
\fi
\epsscale{1.0}
\caption{The S\'{e}rsic indices of galaxies in the CANDELS imaging are plotted vs
  their clustercentric distances. Red stars and black crosses
  represent galaxies classified as ``quiescent'' and ``star-forming'' respectively
  using a rest frame $UVJ$ color selection described in
  \S~\ref{section:cccut}. The open ovals are median values for
  only the quiescent galaxies
  with vertical bars depicting the interquartile range. The
  dashed vertical lines indicate the locations of the cluster core, the
  $1-1.5$ Mpc annulus, and the field. All galaxies above our
  stellar mass limit are plotted in the top panel while
  only intermediate mass galaxies are plotted on the bottom. This plot nicely
  illustrates the drop in S\'{e'}rsic indices for quiescent galaxies
  located within the $1-1.5$ Mpc annulus around the cluster core. It
  can also be seen that only in this region do we find quiescent
  galaxies with S\'{e}rsic indices less than 1 (shown as the horizontal
  dotted line).}\label{fig:serrad}
\ifsubmode
\end{figure}
\else
\end{figure*}
\fi

\ifsubmode
\begin{figure}
\epsscale{1.0}
\else
\begin{figure}[t]
\epsscale{1.15}
\fi
\plotone{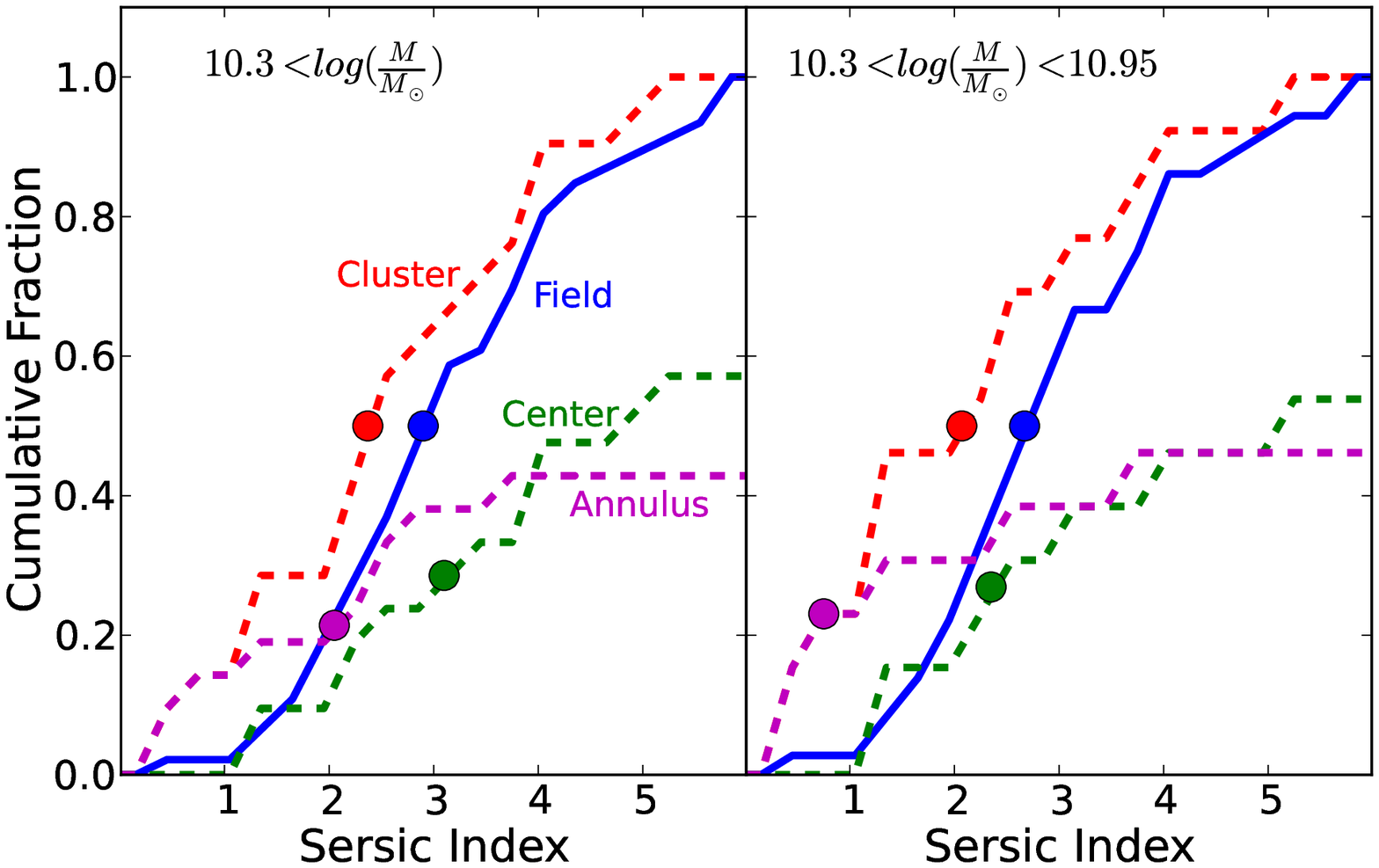}
\ifsubmode
\else
\fi
\epsscale{1.0}
\caption{The cumulative \sersic-index distribution for quiescent
galaxies in the field (solid blue line) and in the cluster (dashed red
line).  The curves for the total cluster and total field samples are 
normalized to the number of galaxies in
each sample. The curves for the cluster subsamples (``field'' and
``annulus'') are normalized by the number of galaxies in the total
cluster sample. The left panel shows the distribution for all quiescent
galaxies with stellar masses $>2 \times 10^{10}$~\msol.  The left
panel shows a subset with moderate stellar mass $2\times 10^{10} - 9
\times 10^{10}$~\msol.    The circular points on each curve show the
median value of the distributions.   The quiescent galaxies associated
with the cluster have smaller median \sersic\ indices, and this is
most pronounced in the moderate mass subsample.    The green (purple)
dashed curve shows the contributions to the cluster cumulative distribution from
quiescent cluster galaxies in the cluster ``center'' with $R < 1$~Mpc
(``annulus'' with 1 Mpc $< R < $1.5 Mpc).
In both panels, the lower \sersic\ indices of  quiescent cluster
galaxies is driven by the quiescent galaxies associated with the
cluster in an  annulus 1 Mpc $< R <$1.5 Mpc, which have the lower
median values.} \label{fig:cumsersic}
\ifsubmode
\end{figure}
\else
\end{figure}
\fi

\subsubsection{Comparison Between the Cluster Core and the\\ 1 Mpc $< \rproj
  <$ 1.5 Mpc Annulus}

As shown in Figure~\ref{fig:cumsersic}, the quiescent galaxies in the
1 Mpc $< \rproj < $ 1.5 annulus  have lower median \sersic\ indices
$n^\mathrm{1-1.5\, Mpc}_\mathrm{med} = 2.1$, compared to
those within 1 Mpc of the center of the cluster, which
have median $n^\mathrm{<1\, Mpc}_\mathrm{med} = 3.0$.   Indeed, four
out of the six moderate mass quiescent galaxies in the
annulus have \sersic\ indices $n < 2.5$, including three galaxies
with $n < 1$.  This is evident in
Figures~\ref{fig:radec} and ~\ref{fig:serrad}, which show that these low-\sersic-index quiescent
galaxies are found preferentially in at 1--1.5~Mpc from the
cluster.  This difference is even more apparent considering the most
restrictive moderate-mass subsample with $2\times 10^{10} - 9\times
10^{10}$~\msol. For this sample, the median \sersic\ indices are
$n=1.0$ and $n=2.4$ for the annulus and cluster core respectively. 

The significance of this result is somewhat limited by the small
number of objects in the samples: there are only 95 quiescent galaxies
in our cluster and field samples combined. To measure the significance
of this result, we considered two Monte Carlo bootstrap simulations to
determine how often we would obtain this result by chance.

We first calculate how often we would obtain the results by chance if
the quiescent galaxies in the 1-1.5 Mpc annulus are randomly drawn
from the parent field sample.   For this test we constructed two
random samples, one having N(1-1.5Mpc) members (the number of
quiescent galaxies in the 1-1.5 Mpc annulus) and the other having
N(field) members. Both samples are drawn from the quiescent field population
with replacement.  From this bootstrap test, we found that the WMW p
value was smaller than the measured value in only 0.44\% of the cases, which
corresponds to a significance of $\approx$ 2.6$\sigma$ (assuming a
Gaussian distribution). Therefore, it seems that even with the small
sample sizes here, the lower \sersic\ indices of the quiescent
galaxies in the annulus 1 Mpc $< \rproj <$ 1.5 Mpc are statistically
significant compared to quiescent galaxies in the field.

We next repeated this bootstrap test to estimate how often we would
obtain the results by chance if the quiescent galaxies in the annulus
are randomly drawn from the parent cluster quiescent sample.  Again,
we constructed random samples, one having N(1-1.5 Mpc) members, and
one having N(<1 Mpc) members.  Here, we find that the likelihood
that the quiescent galaxies in this annulus are drawn from the same
population as those in the rest of the cluster is
$p_\mathrm{bootstrap} = 0.039$ ($\approx 1.8\sigma$) primarily because
of the smaller number of quiescent galaxies in the cluster sample.
Larger samples of quiescent galaxies (especially those in clusters) at
this redshift are required to improve the statistical significance of
this observation.

\ifsubmode
\begin{figure}
\epsscale{1.0}
\else
\begin{figure}[t]
\epsscale{1.15}
\fi
\plotone{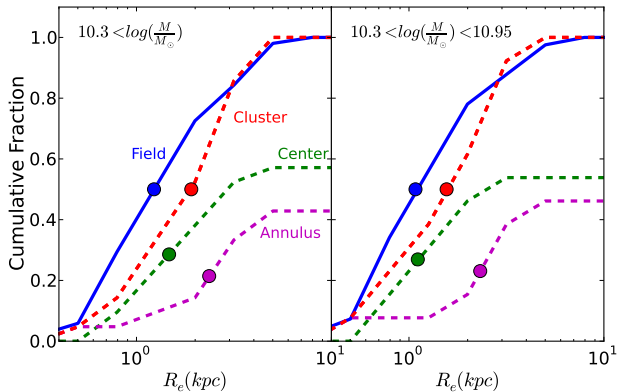}
\ifsubmode
\else
\fi
\epsscale{1.0}
\caption{The cumulative distribution of effective radii for quiescent
galaxies in the field ($\rproj > 3$~Mpc, solid blue line) and in the
cluster ($\rproj < 1.5$~Mpc, dashed red line).  Each curve is
normalized to the total number of galaxies in each sample.  The left
panel shows the distribution for all quiescent galaxies with stellar
masses $>2 \times 10^{10}$~\msol.  The right panel shows a subset with
moderate stellar mass $2\times 10^{10} - 9 \times 10^{10}$~\msol.  The
quiescent galaxies associated with the cluster generally have larger
median effective radii.  In each panel the green (purple) dashed curve
shows the contribution to the cluster cumulative distribution from
quiescent galaxies in the cluster ``center''  with $R < 1$~Mpc
(``annulus'' with 1 Mpc $< R < $1.5 Mpc).  The circle points show the
median value for each distribution.  Quiescent galaxies associated
with the cluster in an annulus 1 Mpc $< R <$1.5 Mpc have larger median
effective radii by factors of 2 compared to the
field.} \label{fig:cumsize}
\ifsubmode
\end{figure}
\else
\end{figure}
\fi

To summarize this section, there is evidence that the
quiescent galaxies in the cluster have lower average  \sersic\ indices
compared to quiescent galaxies in the field.   Furthermore, there is
evidence that this result is driven by a population of quiescent
galaxies in an annulus 1--1.5 Mpc from the cluster. Members of this
population are found to have \sersic\
indices inconsistent with random samples of quiescent galaxies taken from
the cluster core ($\rproj < 1$ Mpc) or from the field. Instead, they
are morphologically more similar to star-forming galaxies found in all
environments.

\subsection{Do the Effective Radii of Galaxies\\Depend on
Environment?}\label{section:sizes}

Because the quiescent galaxies in the cluster show evidence for
smaller \sersic\ indices, we also compared the effective radii of
these galaxies to those in the field.    The median effective radius
of quiescent galaxies with stellar mass $>2 \times 10^{10}$~\msol\
associated with the cluster is $r^\mathrm{cluster}_\mathrm{eff,med} =
2.0$~kpc compared to those in the field which have
$r^\mathrm{field}_\mathrm{eff, med} = 1.4$~kpc.  (We remind the reader
that these radii are the circularized effective radius).  The WMW test
gives a probability, $p = 0.057$.   This is nearly
identical to the findings of \citet{papo12}, who used the same data,
but with slightly different selection criteria and mass limits to
focus on the quiescent galaxies (see also \S~\ref{section:data}).
Figure~\ref{fig:cumsize} shows the cumulative distribution of the
effective radii of quiescent galaxies in the field and cluster, which
illustrates that the cluster galaxy sizes are larger on average.   

We also considered the effective sizes of the quiescent galaxies
within $\rproj < 1$~Mpc of the cluster center compared to those in the
annulus 1 Mpc $< \rproj <$ 1.5 Mpc because these show evidence for
having different \sersic\ indices.  Figure~\ref{fig:cumsize} shows the
cumulative distributions of these subsamples.  The quiescent galaxies
in the 1 Mpc $< \rproj < $ 1.5 Mpc annulus have median
effective radii, $r^\mathrm{1-1.5\, Mpc}_\mathrm{eff, med}=2.7$~kpc,
about 1.3 times larger than other quiescent galaxies in the
cluster or the field.   

Because of the relatively small size of our galaxy samples, we again used a
Monte Carlo bootstrap simulation (similar to the previously described simulations) to test the  significance of these results.
Based on this test, the likelihood that the
quiescent galaxies in this annulus are drawn from the cluster sample is 
9.6\%.  Similarly, the likelihood that the
quiescent galaxies in this annulus are drawn from the same population
as those in the field is 8.3\%.  
Given the relatively small sample size,
it seems that larger samples would be able to further test these
trends. \\
\\
The results of this section can be summarized as follows: we find that
quiescent galaxies associated with the cluster have low S\'{e}rsic
indices and possibly larger effective radii (when compared with quiescent galaxies in the field)
consistent with disk-like morphologies ($n\sim1$). This effect is
driven by galaxies located in an annulus $1-1.5$ Mpc from the center
of the cluster, whereas quiescent galaxies in the cluster core and field
exhibit S\'{e}rsic indices more typical of a bulge dominated
morphology. Furthermore, the most massive galaxies are located in the
cluster core while stellar masses of galaxies in the annulus are
comparable to those found in the field. This suggests that the sample
of quiescent galaxies found $1-1.5$ Mpc from the center of the cluster
is ``contaminated'' with recently quenched star-forming galaxies
accreted from the field.

\section{Discussion}

\subsection{The Relation Between Color, Stellar Mass, and Morphology
  in the Cluster and Field at $z=1.6$} 

We study how the relationship between galaxy color, stellar mass,
and morphology depend on galaxy environment by examining these
relations in the field and in a forming cluster at $z\sim 1.6$.
We find evidence of a possible
correlation between galaxy density with color and mass for 
galaxies with stellar masses $> 2 \times 10^{10}$~\msol\
: compared to galaxies in the field, the cluster galaxies (with
$R_\mathrm{proj} < 1.5$ Mpc) have redder rest frame $U-B$ colors and
higher stellar masses. The color difference appears driven by star-forming galaxies
(we find no difference in the rest-frame $U-B$ colors of quiescent
galaxies).    We interpret this as evidence for increased
dust-obscured star-formation in galaxies closer to the 
cluster, supported by the higher IR-luminosity density in these
galaxies found by \citet{tran10}.  The stellar mass difference appears
driven by the more massive quiescent galaxies with $>9 \times
10^{10}$~\msol (see \S~\ref{section:sfrs} and Figure
\ref{fig:serrad}), and we interpret this as evidence that the most
massive galaxies at this redshift in this field are associated with
the cluster. 

There are morphological differences in the quiescent galaxy
populations associated with the cluster and the field.  Combining our
observations with the results from our previous study \citep{papo12},
we find two main differences between the morphologies of quiescent
galaxies in the field and that of the cluster.  First, \citet{papo12}
found a lack of ``compact'' quiescent galaxies 
with $\reff \sim 1$~kpc in the cluster while such objects are present in the field.
Similarly, many of these quiescent galaxies in the cluster show
evidence for extended disks.    Our analysis here reproduces
the findings of \citet{papo12}, and here we report evidence that the
\sersic\ indices of the quiescent galaxies in the cluster are lower
than for such galaxies in the field.   

\subsection{The Nature of Quiescent Galaxies in the Annulus $\rproj =
  1 - 1.5$ Mpc}

The population of quiescent galaxies in the annulus defined by
projected distances 1 Mpc $< \rproj <$ 1.5 Mpc from the cluster show
the most differences in morphology compared to the field. As discussed
in \S~\ref{section:sersic}, galaxies in this annulus have higher \reff\ and lower
\sersic\ indices.   Indeed,  considering the full subsample with $M >
2\times 10^9$ \msol, their median values are $r_{\mathrm{1-1.5 Mpc}}$
= 2.7 kpc and $n_{\mathrm{1-1.5 Mpc}}$ = 2.1 (compared to medians
$r_{<1\ \mathrm{Mpc}}$ = 1.5 kpc and $n_{<1\ \mathrm{Mpc}}$ = 3.1 in
the cluster center). The quiescent galaxies in the $\rproj$= 1 -- 1.5
Mpc annulus are more typical of star-forming galaxies in the field and
cluster. Figure \ref{fig:stamps} shows examples of these quiescent galaxies in the
cluster annulus with low \sersic\ index ($n < 2$), and compares these
with examples of star-forming galaxies with similar masses and
\sersic\ indices ($n <2$). These examples illustrate that the quiescent galaxies have
morphologies similar to some star-forming galaxies.  In contrast, the
figure also shows quiescent galaxies in the cluster core with more typical,
higher \sersic\ indices ($n > 2$). These have very different morphologies
when compared to annulus quiescent galaxies with low \sersic\ indices. The
star-forming galaxies have an interquartile range of 2.2
- 4.7 kpc in effective radius and 0.8 - 3.3 in \sersic\ index,
consistent with morphologies of these quiescent galaxies in the
annulus. Because these galaxies appear morphologically distinct, we
speculate about their origin and fate.

\ifsubmode
\begin{figure}
\epsscale{1.0}
\else
\begin{figure}[t]
\epsscale{1.15}
\fi
\plotone{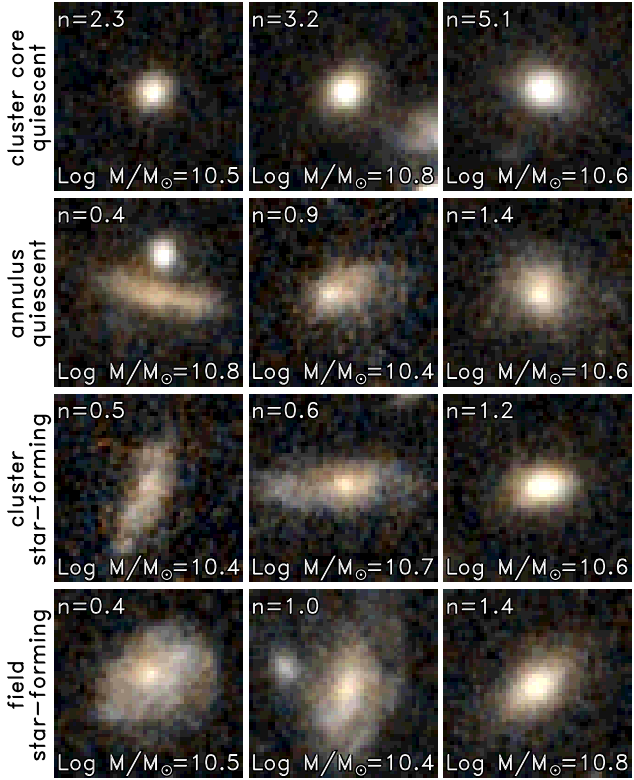}
\ifsubmode
\else
\fi
\epsscale{1.0}
\caption{Examples of quiescent galaxies in the annulus 1 Mpc $< \rproj
  <$ 1.5 Mpc and low \sersic\ index ($n < 2$) and other galaxy
  subsamples. Each panel spans 12\arcsec\ on a side (about 100 kpc at
  $z$=1.6), and the colors show the \hst\ F125W, (F125W+F160W)/2, and
  F160W as blue, green, and red, respectively.     These images
  illustrate that these quiescent galaxies have similar morphologies
  as star-forming galaxies in both the cluster and field selected with
  low \sersic\ indices and over a similar range of stellar mass.   The
  top row shows quiescent galaxies in the core with similar stellar
  masses but more typical, higher \sersic\ indices.} \label{fig:stamps}
\ifsubmode
\end{figure}
\else
\end{figure}
\fi

\ifsubmode
\begin{figure}
\epsscale{1.0}
\else
\begin{figure}[t]
\epsscale{1.15}
\fi
\plotone{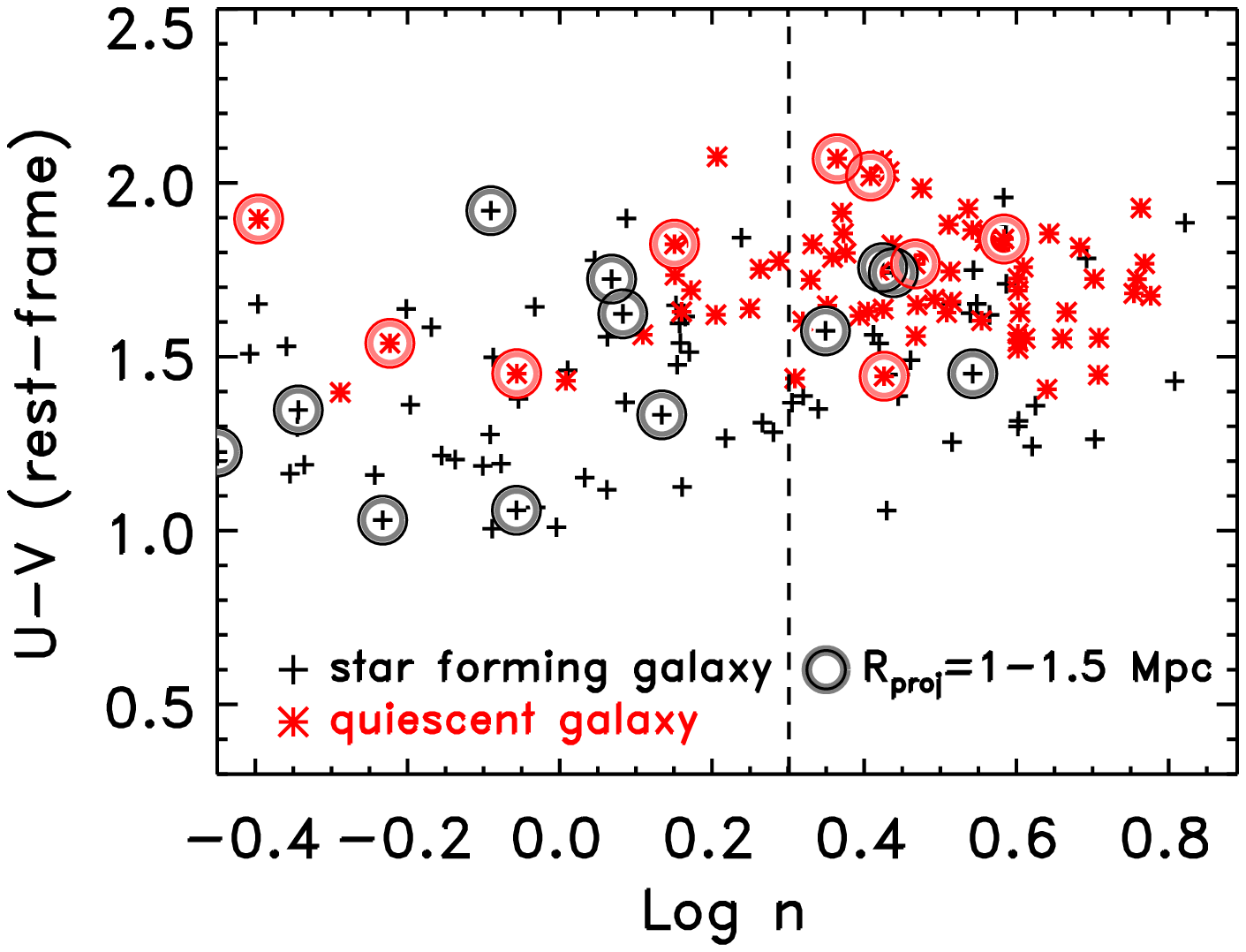}
\ifsubmode
\else
\fi
\epsscale{1.0}
\caption{The \sersic\ indices, $n$, versus the rest-frame $U -V$ color
  for galaxies with stellar mass $>2 \times 10^{10}$~\msol.
  \citet{bell12} have argued that \sersic\ index best correlates with
  increasing $U-V$ color.  We reproduce that result here, as most
  quiescent galaxies lie above $n > 2$, indicated by the dashed line.
      Quiescent and star-forming galaxies in the subsample with 1
Mpc $< R < $1.5 Mpc of the cluster center are indicated with large
circles.  A high fraction ($\approx$50\%) of the quiescent galaxies in
this annulus have $n < 2$, compared to $<$15\% of quiescent galaxies
overall. } \label{fig:umvvn}
\ifsubmode
\end{figure}
\else
\end{figure}
\fi

\citet{bell12} have argued that the \sersic\ index correlates the
strongest with increasing rest-frame $U-V$ color.  We reproduce this
relation in Figure~\ref{fig:umvvn}, which shows the \sersic\ index
versus $U-V$ color for all $z\sim 1.6$ galaxies in our CANDELS data
with stellar mass $> 2 \times 10^{10}$~\msol.   

    We find a similar result to that of Bell et al., as most quiescent
galaxies lie above $n > 2$.  However, many of the quiescent galaxies
associated with the 1 Mpc $<\rproj <$ 1.5 Mpc annulus appear as
outliers.   A high fraction ($\approx$50\%) of the quiescent galaxies
in the annulus 1 Mpc $< \rproj <$ 1.5 Mpc have $n < 2$, compared to
$<$15\% of quiescent galaxies overall.  This 
%
%
leads to the 
%
%
differences in the \sersic\ index distributions we see for quiescent
galaxies in this annulus compared to the other samples
(see \S~\ref{section:sersic}).    

\subsection{Environmental Effects on Galaxy Evolution at $z=1.6$}

At $z=1.6$ there appear to be two evolutionary channels for quiescent
galaxies, one that depends on the mass of the galaxy, and another that depends on
environment. 
Mass is likely related to one channel of quiescence because both the field and cluster 
samples exhibit passive,
bulge dominated galaxies ($n > 2$) 
which show no differences in their
specific SFRs, or rest-frame $U-B$ colors (see
Table~\ref{table}). This mass related channel therefore has only a weak dependence
on environment (cluster versus field) and is likely driven by
processes related to the galaxy halo mass.     This is similar to the
findings of \citet{peng10}. 

The environment also has some role in the morphological evolution of
galaxies.  One difference between the cluster and field quiescent
galaxies is in their sizes. Cluster galaxies have slightly larger effective radii 
than quiescent galaxies in the field \citep[see also][]{papo12}  
significant at $\sim 1.5 \sigma$.
There are also indications that the cluster galaxies have a higher ``dry''
merger rate \citep{lotz11}, which has been interpreted as evidence
that the cluster galaxies have an increased or accelerated
evolution. This view is consistent with the evolution of sizes of
cluster galaxies, and the cluster luminosity function
\citep{papo12,rudn12}.   The higher density region associated with
the cluster appears to increase the merger rate of quiescent galaxies
over that expected solely in the field. In the case
of minor mergers, this has the effect
of increasing their radii with only modestly increasing their
mass. This process appears to work within $\rproj \lsim 1-1.5$ Mpc from
the cluster center.

The environment also seems to affect quiescent galaxies at higher projected
distances from the cluster (1 Mpc $< \rproj <$ 1.5 Mpc).    It is in
this annulus where we see indications that the quiescent galaxies have
larger effective radii and lower \sersic\ indices (see
\S~\ref{section:sersic} and Figure~\ref{fig:serrad}).    The mass
distribution of the quiescent galaxies in this annulus is also
different from quiescent galaxies in the cluster center, which
suggests some process affecting galaxies accreted by the cluster
is linked to the formation of these galaxies. 

As discussed above, the morphologies of the quiescent galaxies in this
annulus are more typical of the morphologies of star-forming,
disk-dominated galaxies.  Interestingly, the projected distances of
the galaxies in this annulus lie outside the upper limits of the
virial radius estimated from galaxy velocity
dispersions \citep[$\simeq 0.9$~Mpc, under the assumption the galaxies
are fully viralized,][]{papo10a} and analysis of the
X-ray data \citep[$\simeq 0.5$~Mpc]{pierre11}. If our findings are
confirmed by larger samples to be a general property of cluster
galaxies at this epoch, it is implied that there is a mechanism
which induces quiescence at large distances from the cluster center
but leaves galaxy morphologies unchanged.

One explanation for these observations is that these disk-dominated
quiescent galaxies at projected distances 1 Mpc $< \rproj <$ 1.5 Mpc
have been preprocessed in smaller group-sized halos, which are now
merging with the cluster \citep[\eg,][]{mcgee09}.
Figure~\ref{fig:radec} shows that the late-type quiescent galaxies may
be located in $\sim$2 groupings. Moreover, qualitatively many of
the $1 < n < 2.5$ quiescent galaxies in the \textit{field} are not
randomly distributed, but rather exist in small groups 
which are identified using Nth nearest neighbor
  measures. \footnote{Characterizing overdensity in our sample using the $N$th
nearest-neighbor method, we 
measured no appreciable differences in galaxy
properties, using either the $z=1.6$ sample defined by $\mathcal{P}_z > 0.4$ or the sample selected using the best-fit photometric redshifts in the range $1.5 < z < 1.75$. This may be due to the fact that our samples rely  largely on photometric redshifts, which is necessary to be complete for red galaxies at $z\sim 1.6$.
Samples based on photometric redshifts  will suffer from projection
effects along the line of sight (even the ``good'' photometric
redshifts of our data, $\Delta z / (1+z) = 0.04$, yield line-of-sight
proper distance uncertainties of 27 Mpc).  This will smear any real
correlations between galaxy density (environment) and galaxy
properties. This is mitigated in lower redshift samples based on
spectroscopic redshifts
\citep[\eg,][]{kauf04,vanderwel08,peng10,cooper11}.  We speculate that
the reason we are able to detect differences in the galaxy properties comparing
the cluster to the field is that the overdensity of the cluster
overcomes the smearing along the line-of-sight from photometric
redshifts.    Nevertheless,  this smearing will affect our results, reducing
any intrinsic trends between galaxy properties and environment.
Therefore, it also follows that the results here show the
\textit{minimum} effect at $z=1.6$ of environment on galaxy properties
such as color, masses, SFRs, 
specific SFRs, sizes, and \sersic\ index. }

However the idea of preprocessing does not
explain why the morphological properties of the quiescent galaxies at
1 Mpc $< \rproj <$ 1.5 Mpc would be significantly different from
quiescent field galaxies. 
We favor an explanation where quiescent galaxies at projected
distances of 1 Mpc $< \rproj <$ 1.5 Mpc began as star-forming,
disk-dominated galaxies before they entered the proximity of the
forming galaxy cluster.  At these distances they undergo truncation of
their star formation without disrupting their morphological profiles
or sizes.   At later times they will enter the cluster ``halo'',  and,
because they are quiescent, likely will undergo dissipationless
mergers with other quiescent galaxies.  This will then reconfigure
their morphologies, consistent with the enhanced merger rate of
\citet{lotz11}.    The mechanism for this transition from star-forming
disk galaxy to quiescent galaxy appears to begin outside the cluster
virial radius. If this occurs in galaxy groups, then it affects the
galaxies only as they approach the region of the cluster. 

This interpretation may be explained by the expectation from theory
that as gas falls into the region of large halos it is shocked to the
virial temperature of the halo \citep[\eg,][]{white78,birn03}.
Recent cosmological $N$-body and hydrodynamical simulations show that
the velocity shock from the virialized region extends beyond the
virial radius \citep{cuesta08,birn07,dekel09a}, and this has effects
on both the baryonic gas fraction, and the velocities and temperatures
of the ``shocked'' infalling gas out to $1.5-2 R_\mathrm{vir}$ as well
as the stripping of subhalos at $3-4 R_\mathrm{vir}$.
\citep[E. Zinger, A. Dekel, et al., in preparation; P. Behroozi, R. Wechsler, et al. in preparation]{krav05,bahe12}.
 
The quiescent disk-dominated galaxies we observe at $1-1.5$ Mpc from
the cluster may be inside this hot (shocked) medium, where gas
accretion to them is then shut off \citep{dekel06a,croton06}.  This
process is  often referred to as gas ``strangulation''
\citep{balogh00}.  It is possible these galaxies are on their first
infall into the cluster: the cluster-crossing time is 5 Gyr (assuming
the 1 Mpc radius and typical peculiar velocity of $\approx$400 km
s$^{-1}$, Pierre et al.\ 2012), and this exceeds the age of the
Universe at $z=1.6$.  Under the assumption that the galaxies are on
their first pass through the cluster (and have not already passed
through), effects from a virial-shock-like process seem favored
because the galaxies lie outside the central cluster region.  Other
processes that could affect these galaxies such as ram-pressure
stripping \citep{gunn72,abadi99}, tidal stripping \citep{read06}, and
galaxy harassment \citep{moore96} are expected to be more significant
nearer the central cluster regions where the galaxy and gas densities
are higher.  Because it is less likely that the quiescent
disk-dominated galaxies at $1-1.5$~Mpc have had sufficient time to
pass through the cluster,  it seems more likely that they have
consumed their gas supply and gas accretion has been ``strangulated''
by the hot medium. 

We also cannot rule out the possibility, however, that these galaxies
\textit{have} already passed through the outskirts of the larger dark matter
potential and are now located at the turn around radius of the
cluster. By passing through the central halo at large radii, the
baryonic content of this population could have been altered without
significantly restructuring their morphologies. Although we estimate
that the crossing times for this cluster may be longer than the age of
the universe at this redshift, this structure has grown since earlier
epochs, and as it does not yet appear virialized the interpretation of
the kinematics is unclear.
 
Another mechanism that may explain this processing of galaxies at large
clustercentric distances is the assembly bias of dark matter halos
\citep{gao05,wech06,croton07}.   Simulations show that  halos with early
formation redshifts are more strongly clustered than halos of
comparable mass, which form later.  Assembly bias in the nearby
universe $(0.01<z<0.3)$ has been studied in the SDSS by
\citet{cooper09} who find that galaxies with older stellar populations
and higher stellar metalicities are preferentially found in higher
density environments. This suggests that, at a given mass, early type
galaxies in dense environments were formed earlier, and thus ceased
forming stars at an earlier epoch. This idea is supported by
the comparison of ages of low redshift early-type galaxies
between low and high local number density regions. Massive galaxies in
high local number density regions appear to form their stellar populations in the range 
between $z\sim 2$ and $z\sim 5$, while their low local number density
counterparts are $1-2$ Gyr younger \citep{thomas05}.

Simulations show that a cause for this assembly bias is that
group-sized halos must compete for mass (both baryonic and dark matter) when in close proximity to
much larger dark matter halos.  In effect, the group halos will be
starved of dark matter and the galaxies they contain will be similarly
starved of baryonic gas. These galaxies will exhaust their supply of
star-forming gas more quickly than galaxies located in similar halos
in isolation. As this starvation is not directly associated with
mergers or other more direct galaxy interactions, it is likely that
the cessation of star formation will not disrupt a galaxy's late-type
morphology.  Our observations at $z=1.6$ may be due to this same
affect: differences we see between field galaxies and those in the
outskirts of the cluster may be due to an earlier formation of dark
matter halos in regions of high density.  
\citet{bell12} studied galaxies over a large range of redshift in
the CANDELS UDS field, finding the $z\sim 1.6$ redshift slice 
to contain a high local number
density of galaxies (presumably many groups). 
There are more quiescent galaxies with low \sersic\ indices
in this redshift slice than are seen in other slices. This reinforces the notion that
quiescent galaxies with low \sersic\ indices are the result of
environment processes. 

If the environment of the cluster does suppress gas accretion onto
galaxies, then we naturally expect a higher quiescent fraction of
galaxies in the vicinity of the cluster.  We test this using the
fraction of quiescent galaxies, $f_Q$, which is the ratio of
the number of quiescent galaxies to the total number of galaxies.
Taking galaxies with stellar mass $> 2 \times 10^{10}$~\msol\ from the
full UDS sample at the redshift of the cluster, we estimate an intrinsic quiescent fraction of $f_Q =
0.53$. This is consistent with the measurements from \citet{quadri12}
considering differences in redshift binning, and in the
measurements of the internal colors \citep[see discussion in \S~\ref{section:data},][]{quadri12,bram09}. 

\ifsubmode
\begin{figure}
\epsscale{1.0}
\else
\begin{figure}[t]
\epsscale{1.15}
\fi
\plotone{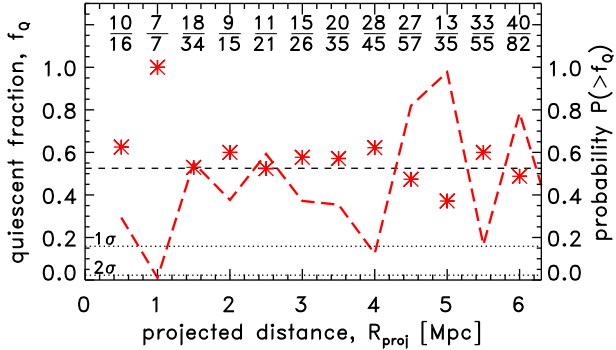}
\ifsubmode
\else
\fi
\epsscale{1.0}
\caption{The fraction of quiescent galaxies, $f_Q$,  as a function of
projected distance from the center of the cluster.  Galaxies
have stellar mass $> 2\times 10^{10}$~\msol.   Each datum shows the
ratio of the number of quiescent galaxies to the total in each bin of
projected distance.  This fraction is printed above each datum.  The
horizontal dashed line shows the quiescent fraction of all galaxies in
the same (including both cluster and field samples), $f_Q = 0.52$.
The long-dashed curve shows the probability of obtaining the measured
fraction or greater assuming a binomial distribution assuming an
intrinsic fraction of $f_Q$.    The dotted lines show significance
levels of 1 and 2$\sigma$.    The binomial probability is very
dependent on the intrinsic fraction of $f_Q$, and the value we assume
is a conservative upper limit for reasons in the text.  At projected
distances $\rproj < 1$~Mpc, we observe a significant enhancement of the
quiescent fraction, but this does not  extend beyond 1
Mpc. } \label{fig:qfrac}
\ifsubmode
\end{figure}
\else
\end{figure}
\fi
 
Figure~\ref{fig:qfrac} shows the $f_Q$  as a function of projected
distance from the center of the cluster.   The figure also
shows the probability of obtaining the expected $f_Q$ value for a binomial
distribution given the numbers of galaxies in each bin.
Galaxies within 1 Mpc show an enhancement in the
quiescent fraction, with $f_Q$=0.62-1.0.  Using a binomial distribution,
this is significant with $p=0.033$ for all galaxies within 1 Mpc,
about the 2$\sigma$ level assuming a Gaussian distribution (and even
higher significance, $p=0.011$, considering the galaxies at 0.5--1 Mpc
only).  

There is evidence for an enhanced
quiescent fraction in the cluster.   At projected distances
$\rproj < 1$~Mpc, we observe a significant enhancement ($> 2 \sigma$)
of the quiescent fraction (Figure~\ref{fig:qfrac}).  
In the annulus 1 Mpc $< \rproj <$ 1.5 Mpc, the galaxies have a
quiescent fraction that is consistent with the intrinsic fraction
derived for all galaxies.  As illustrated in Figure \ref{fig:serrad} the unchanging
quiescent fraction from the annulus to the field results of a high
number of  star-forming galaxies in the annulus.   We are unable to
rule out the possibility that these star-forming galaxies result from
contamination effects, and that the intrinsic quiescent fraction in
the annulus is higher.   It may also be possible that the some
processes acting to suppress star-formation do not follow a
spherically symmetric geometry (which we assume using an annulus of
constant projected radius).  Larger samples of cluster galaxies
observed at large projected distances are needed to make more rigorous
conclusions.   Nevertheless, the evidence here is that  the galaxies
in the 1 Mpc $< \rproj <$1.5 Mpc annulus have a quiescent fraction
consistent with the field.  This evidence disfavors (but does not
exclude) the assembly bias scenario as this effect
would likely exhibit an enhanced quiescent fraction in the near vicinity of
the assembling cluster. 

Our findings at $z=1.6$ are similar to some studies at lower
redshifts.  \citet{woo12} show that at $z\sim 0-0.2$ and fixed stellar
mass, the fraction of quiescent satellite galaxies is strongly dependent
on the projected distance from the center of the parent halo, and that
the fraction of ``quenched'' galaxies is enhanced out to projected
distances of $\simeq 1.5 R_\mathrm{vir}$ around massive halos ($M_h
\sim 10^{14.5}$~\msol).  \citet{wein06a} reach similar
conclusions. \citet{wetzel11b} similarly find that the fraction of
quenched galaxies that are centrals in their halos is higher and
deviates significantly from field values  within $2\times
R_\mathrm{vir}$, but at larger clustercentric distances there is no
indication that the environment has any effect on the star-formation
in galaxies.  Depending on the size of the virial radius of the
cluster (upper limit of $R_\mathrm{vir} \simeq 0.9$ Mpc, see
above), it appears that the quiescent fraction is enhanced out to
distances of $\approx 1 R_\mathrm{vir}$.  Interestingly, the enhanced
quiescent fraction does not extend to projected distances of
$1-1.5$~Mpc, which would be expected if field galaxies are being
quenched as they are accreted into the cluster environment. A possible 
explanation is that the galaxies in the annulus are in a group sized
dark matter halo(s), and may experience excess star-formation
\citep[\eg,][]{tran09} in addition to quenching that keeps the
quiescent fraction approximately equivalent to the field value.

Another interesting system is the well
studied, X-ray luminous cluster XMMU J2235.3-2557 at a redshift of $z
= 1.39$ \citep{lidman08,strazz10,bauer10}. The core of this highly
evolved cluster exhibits a tight red sequence of galaxies (very
similar to the cluster at $z=1.62$ in our field) with little
or no ongoing star formation. The cluster members on the outskirts on
the other hand show greater diversity and a notable increase in [OII]
emitters and galaxies, which may be hosts of dust obscured star
formation. There is also evidence for a ``quenching radius'' (located
$\sim 200$ kpc from the cluster core) within which star formation is
rapidly truncated, possibly in relation to the hot, X-ray emitting
intracluster medium.  \citet{grutz12} extend the study out to $\sim 1.5 R_{vir}$ and
find SFRs lower than typical values for field galaxies at the same
redshift.  It is noted that clusters detected due to their extended
X-ray  emission are more likely to be evolved structures than clusters
detected by different methods.  Observations of XMMU J2235.3-2557 are
consistent with more advanced stages of the evolution proposed for the
cluster studied here.

Therefore, our interpretation of the data is that environments of
higher density, such as the forming cluster at $z=1.62$, accelerate
the morphological evolution and quiescent fraction of galaxies.
Furthermore the data show that the environment has two effects.
First, Lotz et al.\ (2013) and Papovich et al.\
(2012) argue for an increased merger rate within the densest
environments in order to account for the larger effective sizes of
quiescent galaxies and the elevated number of faint companions
observed.   Second, here we argue that as galaxies
enter the sphere of influence of the cluster, star-formation must be
suppressed in some galaxies in order to explain their quiescent colors
and morphologies (\S~\ref{section:sersic}).  Theory predicts that the
physical effects associated with the cluster could include gas-shock
heating, as discussed above. Another possibilty that can not be
rejected is that galaxies near the cluster
may experience assembly bias as they compete for baryonic
gas with the larger gravitational potential of the cluster.  Furthermore, \citet{quadri12} observe
an upturn in the fraction of quiescent galaxies for lower-mass
galaxies in higher density regions in the same mass and redshift range where we
observe the change in morphologies, adding further evidence that the
environment has a strong effect on galaxies of moderate mass.
Therefore, at $z=1.6$, it appears that the environment
has a stronger effect on moderate-mass galaxies ($2 \times 10^{10} -
9\times 10^{10}$~\msol), as it is these galaxies that exhibit the most
difference in size and \sersic\ index compared to similar galaxies in
the field \citep[although we note that this applies less to the most
massive quiescent galaxies, which show very similar sizes in both the
cluster and field,][]{papo12}.  

\section{Summary}

To summarize, in this paper we have studied the
dependence of galaxy color, stellar mass, and morphology for galaxies
on environment at $z=1.6$ in the CANDELS/UDS field. This field containts
\textit{Hubble Space Telescope} near-infrared imaging over $9\farcm4 \times 22\farcm0$, corresponding to a projected, physical area of 4.8
Mpc $\times$ 11.2 Mpc at $z=1.6$.  This field also contains a known
galaxy overdensity (cluster) at this redshift \citep{papo10a}.  We
define a sample of cluster galaxies as those within 1.5 Mpc of the
cluster center, and we define a sample of galaxies in the field as
those further than 3.0 Mpc of the cluster center. We also find it
useful to define a sample of galaxies in an annulus between
$\rproj=1-1.5$ Mpc from the cluster center. Because we use
the same suite of ground-based and space--based data to study the
properties of the galaxies in our samples, our study is very
homogenous.  Any systematic biases affecting our data or analysis will
be identical, and studies of the relative properties between the
cluster and field samples will be robust.  

We derive stellar masses, rest-frame colors, and SFRs using the broad
suite of ground-based and \spitzer\ data.   We quantify the morphology
of the galaxies using GALFIT with the \hst/F125W data to measure the
effective radius, $\reff$,  and \sersic\ index, $n$, of all galaxies
in our samples.   In both the cluster and field, half of the
bulge-dominated galaxies ($n>2$) reside on the red sequence
of the color-magnitude diagram, and most
disk-dominated galaxies ($n<2$) reside in the blue cloud associated
with star-forming galaxies.  Applying the WMW statistical test to
field and cluster galaxies, we derive some evidence that the cluster
galaxies have redder rest-frame $U-B$ colors and higher stellar masses
compared to the field.  star-forming galaxies in both the cluster and
field show no significant differences in their morphologies.  In
contrast, there is evidence that
quiescent galaxies in the cluster have larger median effective radii,
$r^\mathrm{cluster}_\mathrm{eff,med}=2.0$~kpc, and smaller \sersic\
indices,  $n^\mathrm{cluster}_\mathrm{med}=2.6$ compared to the field,
which have $r^\mathrm{field}_\mathrm{eff,med}=1.4$~kpc and
$n^\mathrm{field}_\mathrm{med}=3.3$.

The differences in morphology are larger comparing quiescent galaxies
in an annulus defined by clustercentric distances
1~Mpc~$<\rproj<$~1.5~Mpc to the other quiescent galaxies.  The quiescent
galaxies in this annulus have median $n^\mathrm{1-1.5\,
Mpc}_\mathrm{med}=2.1$ and $r^\mathrm{1-1.5\,
Mpc}_\mathrm{reff,med}=2.7$~kpc, more consistent with the morphologies
of the star-forming galaxies. 
We find that the \sersic\ index generally correlates with rest-frame $U-V$
color, such that the majority of red, quiescent galaxies have higher
\sersic\  indices ($n > 2$), consistent with the findings of
\citet{bell12}.    However, a high fraction ($\approx 50$\%) of the
quiescent galaxies in the 1 Mpc $< \rproj <$ 1.5 Mpc annulus have $n <
2$ compared to only $< 15$\% of quiescent galaxies overall. We argue
that these galaxies have been processed under the influence of the
cluster environment, and the evidence favors models where gas
accretion onto these galaxies is suppressed.  

We argue that the primary channel for the evolution of
galaxies at $z=1.6$ depends on the mass of the galaxy, as the galaxies
in both the cluster and field show weak differences in their  masses,
specific SFRs, and rest-frame $U-B$ colors. 

Our results also show that the environment has some role in the
morphological evolution of galaxies at $z=1.6$.   These differences are seen in
the cluster as a lack of 
``compact'' quiescent galaxies observed in the field \citep[see
see also,][]{papo12}.   The quiescent cluster galaxies also have lower
\sersic\ indices compared to the field, and this is especially true
for the quiescent galaxies in the annulus of 1--1.5 Mpc from the
cluster.    The change in galaxy properties in this annulus is
especially interesting as these clustercentric distances are greater
than estimates of the virial radius, which have an upper limit of 0.9
Mpc from dynamical and X-ray constraints (see Papovich et al.\ 2010
and Pierre et al.\ 2011).    Taken together, we argue that the
environment influences star-forming galaxies as they enter the sphere of
influence of the cluster at $\sim$2 times the virial radius, 
and that this may affect more strongly the
moderate mass galaxies.  This is consistent with expectations from
theory, including effects from strangulation, preprocessing by groups,
or assembly bias. Our data favors a scenario dominated by
strangulation, however the other two possibilities cannot be excluded.

As a final word,  it is unknown if our results are extendable to all
quiescent galaxies in different environments at $z=1.6$, including
better statistics on group environments.  Our results
are based on a single forming galaxy cluster at $z\sim 1.6$, and even with
the large CANDELS dataset, we have morphological information from
\hst\ for fewer than 100 quiescent galaxies at this redshift.  Clearly
larger samples of  galaxies at this (and other) redshifts are required
to generalize these results and to improve (or refute) the
significance of the differences in morphology we observe here.  Such
tests will be possible using the full CANDELS dataset as well as other
\hst\ studies in fields of high redshift clusters.   

\acknowledgements

We wish to thank the members of the CANDELS team for their
contributions, and we thank L.~Macri, and R. Quadri  for stimulating
discussions, and helpful comments. We also wish to thank the anonymous
referee for comments and suggestions that improved the quality and
clarity of this work. This work is based on observations
taken by the CANDELS Multi-Cycle Treasury Program with the NASA/ESA
HST, which is operated by the Association of Universities for Research
in Astronomy, Inc., under NASA contract NAS5-26555.  This work is
supported by HST program number GO-12060.  Support for Program number
GO-12060 was provided by NASA through a grant from the Space Telescope
Science Institute, which is operated by the Association of
Universities for Research in Astronomy, Incorporated, under NASA
contract NAS5-26555.   This work is based on observations made with
the \textit{Spitzer Space Telescope}, which is operated by the Jet
Propulsion Laboratory, California Institute of Technology.  This work
is based in part on data obtained as part of the UKIRT Infrared Deep
Sky Survey. MCC acknowledges support provided by NASA through Hubble
Fellowship grant \#HF-51269.01-A, awarded by the Space Telescope
Science Institute, which is operated by the Association of
Universities for Research in Astronomy, Inc., for NASA, under contract
NAS 5-26555. MCC also acknowledges support from the Southern
California Center for Galaxy Evolution, a multi-campus research
program funded by the University of California Office of Research.
DC acknowledges receipt of a QEII Fellowship from the Australian
Government. We acknowledge generous support from the George P. and
Cynthia Woods Institute for Fundamental Physics and Astronomy. 

\bibliographystyle{apj}

\bibliography{apj-jour,alpharefs}{}


\begin{appendix}
\section{GALFIT Tests}
\label{sec:appendix}
In order to determine to which magnitude we can reliably extract
structural parameters of a galaxy using GALFIT, we performed a series
of simulations. Galaxy models were created by running GALFIT with all
fitted parameters (such as effective radius, S\'{e}rsic index, and
magnitude) fixed to chosen values. The models were then inserted
randomly into the HST F125W image, avoiding the edge and extreme
proximity or overlap with existing objects. Then GALFIT was run on
these simulated galaxies embedded in our data in an attempt to recover the known
values. The input models tested span a range of S\'{e}rsic indices,
effective radii, and magnitudes typical of galaxies in our sample. The
output values from these tests were then separated into bins $0.5$ mag
in size and the robust median and biweighted scale \citep[as described
in][]{beers90} were computed for various input parameters. The
results are summarized in table \ref{table2}, and plotted in Figure
\ref{fig:gsims}. As previously stated, our analysis of these
simulations showed that the recovered effective radii are accurate to better than 40\% and the
\Sersic\ indices to better than 25\% for galaxies with $\reff=2$ kpc
and $n=4$ with magnitude $m(F125W)=24$ mag,  near our stellar-mass
limit of our sample, ($2 \times 10^{10}$~\msol). Galaxies of this type
are among the most difficult to fit: the effective radius and
S\'{e}rsic index are highly covariant.
Uncertainties are significantly lower for more disk-like and brighter
sources (as shown in Table~\ref{table2}). 

\ifsubmode
\begin{figure}
\epsscale{1.0}
\else
\begin{figure*}
\epsscale{1.0}
\fi
\plotone{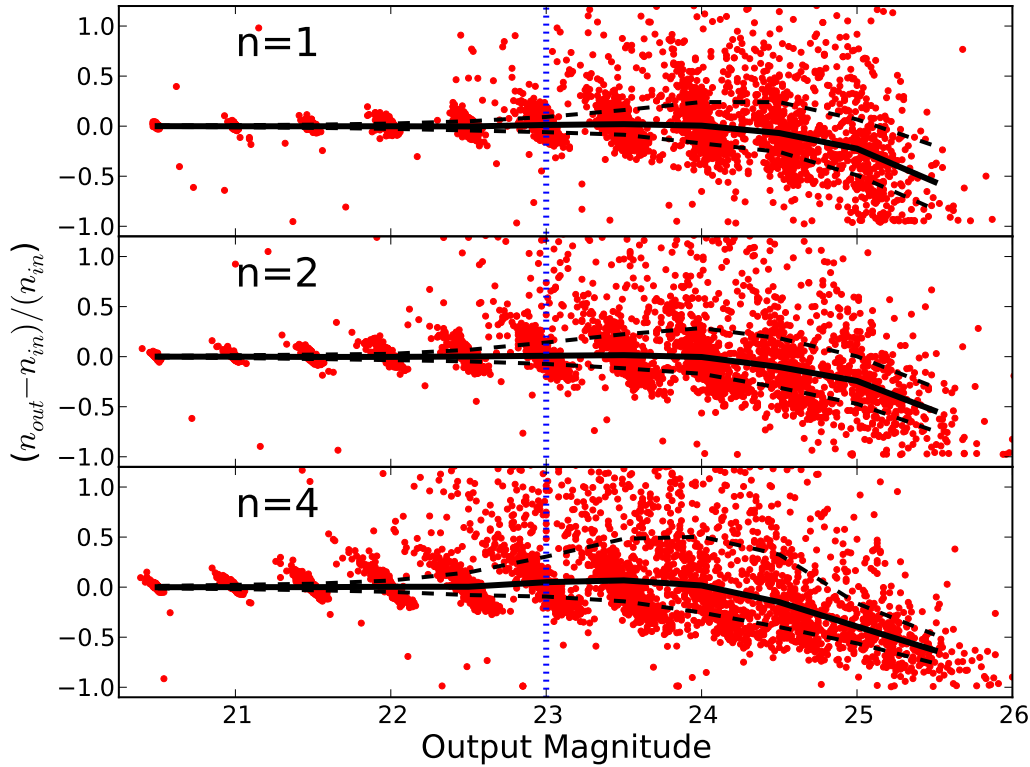}
\epsscale{1.0}
\vspace{12pt}
 \caption{The GALFIT output magnitude vs fractional difference in
   \sersic\ index (left panel) and effective radius (right panel) between input and output
   values. The blue points show the median values, the error bars
   depict the biweighted scale ($S_{BI}$) as described in \citet{beers90}. The points
   are placed in half magnitude wide bins with the center of the first
   bin at 20.5. The panels are separated vertically by \sersic\
   indices and effective radii values (respectively) used as input in
   our simulations. Note that because each panel is marginalized over
   the other parameter (e.g. n=1 plot contains all $r_{eff}$), this
   plot is not directly comparable to Table \ref{table2}. Note also
   that the tendency for datapoints to be arranged in diagonal lines
   is due to the fact that input magnitudes were chosen at 0.5
   magnitude steps. This is similar to the simulations of \citet{haus07}.} \label{fig:gsims}
\ifsubmode
\end{figure}
\else
\end{figure*}
\fi

\clearpage
\LongTables 
\begin{deluxetable}{lcccc}
\tablecaption{Summary of GALFIT tests for HST WFC3 F125W Data\label{table2}}
\tablewidth{0pc}
\tablehead{
\colhead{\textbf{Output Magnitude}} &
   \colhead{\textbf{$r_{out}$ Range (kpc)}} &
   \colhead{\textbf{$n_{out}$ Range}} &
   \colhead{\textbf{$S_{BI}[(r_{out}-r_{in})/r_{in}]$}} &
   \colhead{\textbf{$S_{BI}[(n_{out}-n_{in})/n_{in}]$}}
}

\startdata
20.5$\pm$0.25 & 0-.75 & $<$1 & 0.002 & 0.005\\
& & 1-2.5 & 0.003 & 0.010\\
& & > 2.5 & 0.010 & 0.016\\
& 1.0$\pm$.25 & $<$1 & 0.002 & 0.004\\
& & 1-2.5 & 0.006 & 0.009\\
& & > 2.5 & 0.018 & 0.018\\
& 1.5$\pm$.25 & $<$1 & 0.004 & 0.005\\
& & 1-2.5 & 0.009 & 0.012\\
& & > 2.5 & 0.024 & 0.020\\
& 2.0$\pm$.25 & $<$1 & 0.005 & 0.005\\
& & 1-2.5 & 0.012 & 0.013\\
& & > 2.5 & 0.031 & 0.021\\
\hline\\[-2ex]
21.0$\pm$0.25 & 0-.75 & $<$1 & 0.003 & 0.008\\
& & 1-2.5 & 0.005 & 0.018\\
& & > 2.5 & 0.017 & 0.026\\
& 1.0$\pm$.25 & $<$1 & 0.004 & 0.007\\
& & 1-2.5 & 0.009 & 0.015\\
& & > 2.5 & 0.031 & 0.031\\
& 1.5$\pm$.25 & $<$1 & 0.005 & 0.008\\
& & 1-2.5 & 0.015 & 0.020\\
& & > 2.5 & 0.044 & 0.034\\
& 2.0$\pm$.25 & $<$1 & 0.008 & 0.010\\
& & 1-2.5 & 0.019 & 0.020\\
& & > 2.5 & 0.065 & 0.044\\
\hline\\[-2ex]
21.5$\pm$0.25 & 0-.75 & $<$1 & 0.005 & 0.012\\
& & 1-2.5 & 0.009 & 0.026\\
& & > 2.5 & 0.025 & 0.041\\
& 1.0$\pm$.25 & $<$1 & 0.006 & 0.010\\
& & 1-2.5 & 0.017 & 0.028\\
& & > 2.5 & 0.046 & 0.046\\
& 1.5$\pm$.25 & $<$1 & 0.010 & 0.015\\
& & 1-2.5 & 0.025 & 0.033\\
& & > 2.5 & 0.067 & 0.049\\
& 2.0$\pm$.25 & $<$1 & 0.013 & 0.015\\
& & 1-2.5 & 0.031 & 0.035\\
& & > 2.5 & 0.101 & 0.068\\
\hline\\[-2ex]
22.0$\pm$0.25 & 0-.75 & $<$1 & 0.006 & 0.017\\
& & 1-2.5 & 0.015 & 0.042\\
& & > 2.5 & 0.048 & 0.078\\
& 1.0$\pm$.25 & $<$1 & 0.011 & 0.017\\
& & 1-2.5 & 0.025 & 0.040\\
& & > 2.5 & 0.059 & 0.065\\
& 1.5$\pm$.25 & $<$1 & 0.015 & 0.023\\
& & 1-2.5 & 0.036 & 0.048\\
& & > 2.5 & 0.086 & 0.071\\
& 2.0$\pm$.25 & $<$1 & 0.017 & 0.027\\
& & 1-2.5 & 0.053 & 0.058\\
& & > 2.5 & 0.256 & 0.158\\
\hline\\[-2ex]
22.5$\pm$0.25 & 0-.75 & $<$1 & 0.014 & 0.030\\
& & 1-2.5 & 0.024 & 0.065\\
& & > 2.5 & 0.068 & 0.109\\
& 1.0$\pm$.25 & $<$1 & 0.018 & 0.022\\
& & 1-2.5 & 0.042 & 0.065\\
& & > 2.5 & 0.113 & 0.106\\
& 1.5$\pm$.25 & $<$1 & 0.023 & 0.034\\
& & 1-2.5 & 0.056 & 0.070\\
& & > 2.5 & 0.140 & 0.120\\
& 2.0$\pm$.25 & $<$1 & 0.031 & 0.047\\
& & 1-2.5 & 0.087 & 0.090\\
& & > 2.5 & 1.033 & 0.643\\
\hline\\[-2ex]
23.0$\pm$0.25 & 0-.75 & $<$1 & 0.018 & 0.047\\
& & 1-2.5 & 0.038 & 0.092\\
& & > 2.5 & 0.095 & 0.161\\
& 1.0$\pm$.25 & $<$1 & 0.027 & 0.039\\
& & 1-2.5 & 0.063 & 0.106\\
& & > 2.5 & 0.152 & 0.161\\
& 1.5$\pm$.25 & $<$1 & 0.036 & 0.051\\
& & 1-2.5 & 0.089 & 0.110\\
& & > 2.5 & 0.222 & 0.196\\
& 2.0$\pm$.25 & $<$1 & 0.053 & 0.070\\
& & 1-2.5 & 0.129 & 0.135\\
& & > 2.5 & 3.571 & 0.881\\
\hline\\[-2ex]
23.5$\pm$0.25 & 0-.75 & $<$1 & 0.033 & 0.074\\
& & 1-2.5 & 0.056 & 0.154\\
& & > 2.5 & 0.153 & 0.253\\
& 1.0$\pm$.25 & $<$1 & 0.044 & 0.066\\
& & 1-2.5 & 0.089 & 0.148\\
& & > 2.5 & 0.213 & 0.223\\
& 1.5$\pm$.25 & $<$1 & 0.063 & 0.074\\
& & 1-2.5 & 0.116 & 0.161\\
& & > 2.5 & 0.275 & 0.231\\
& 2.0$\pm$.25 & $<$1 & 0.093 & 0.094\\
& & 1-2.5 & 0.200 & 0.202\\
& & > 2.5 & 3.360 & 1.019\\
\hline\\[-2ex]
24.0$\pm$0.25 & 0-.75 & $<$1 & 0.060 & 0.132\\
& & 1-2.5 & 0.108 & 0.273\\
& & > 2.5 & 0.186 & 0.301\\
& 1.0$\pm$.25 & $<$1 & 0.071 & 0.098\\
& & 1-2.5 & 0.138 & 0.251\\
& & > 2.5 & 0.332 & 0.308\\
& 1.5$\pm$.25 & $<$1 & 0.086 & 0.123\\
& & 1-2.5 & 0.194 & 0.285\\
& & > 2.5 & 0.384 & 0.359\\
& 2.0$\pm$.25 & $<$1 & 0.132 & 0.189\\
& & 1-2.5 & 0.273 & 0.305\\
& & > 2.5 & 3.243 & 0.940\\
\end{deluxetable}

\end{appendix}

\end{document}